\documentclass[english,aps,superscriptaddress]{revtex4}
\usepackage[T1]{fontenc}
\usepackage[latin9]{inputenc}
\usepackage{verbatim}
\usepackage{float}
\usepackage{amsmath}
\usepackage{amssymb}
\usepackage{graphicx}
\usepackage{esint}

\makeatletter

\providecommand{\tabularnewline}{\\}
\newcommand{\lyxdot}{.}

\@ifundefined{textcolor}{}
{%
 \definecolor{BLACK}{gray}{0}
 \definecolor{WHITE}{gray}{1}
 \definecolor{RED}{rgb}{1,0,0}
 \definecolor{GREEN}{rgb}{0,1,0}
 \definecolor{BLUE}{rgb}{0,0,1}
 \definecolor{CYAN}{cmyk}{1,0,0,0}
 \definecolor{MAGENTA}{cmyk}{0,1,0,0}
 \definecolor{YELLOW}{cmyk}{0,0,1,0}
}

\usepackage{babel}

\makeatother

\usepackage{babel}
\begin{document}

\title{Rapidity dependence of global polarization in heavy ion collisions}

\author{Zuo-Tang Liang}

\email{liang@sdu.edu.cn}

\affiliation{Institute of Frontier and Interdisciplinary Science, Key Laboratory
of Particle Physics and Particle Irradiation (MOE), Shandong University,
Qingdao, Shandong 266237, China}

\author{Jun Song}

\email{junsong2006@163.com}

\affiliation{Department of Physics, Jining University, Jining, Shandong 273155,
China}

\author{Isaac Upsal}

\email{iupsalp@gmail.com}

\affiliation{Institute of Frontier and Interdisciplinary Science, Key Laboratory
of Particle Physics and Particle Irradiation (MOE), Shandong University,
Qingdao, Shandong 266237, China}

\affiliation{Physics Department, Brookhaven National Laboratory, Upton, New York
11973, USA}

\author{Qun Wang}

\email{qunwang@ustc.edu.cn}

\affiliation{Department of Modern Physics, University of Science and Technology
of China, Hefei, Anhui 230026, China}

\author{Zhang-Bu Xu}

\email{xzb@bnl.gov}

\affiliation{Institute of Frontier and Interdisciplinary Science, Key Laboratory
of Particle Physics and Particle Irradiation (MOE), Shandong University,
Qingdao, Shandong 266237, China}

\affiliation{Physics Department, Brookhaven National Laboratory, Upton, New York
11973, USA}
\begin{abstract}
We use a geometric model for the hadron polarization with an emphasis
on the rapidity dependence. It is based on the model of Brodsky, Gunion,
and Kuhn and that of the Bjorken scaling. We make predictions for
the rapidity dependence of the hadron polarization in the collision
energy range 7.7-200 GeV by taking a few assumed forms of the parameters.
The predictions can be tested by future experiments. 
\end{abstract}
\maketitle

\section{Introduction}

In non-central collisions of heavy ions at high energies a huge orbital
angular momentum (OAM) is generated. Through spin-orbit couplings
in parton-parton scatterings, hadrons can be globally polarized along
the OAM of two colliding nuclei \cite{Liang:2004ph,Liang:2004xn,Voloshin:2004ha}.
In a hydrodynamic picture, the huge OAM is distributed into a fluid
of quarks and gluons in the form of local vorticity \cite{Betz:2007kg,Becattini:2007sr,Becattini:2015ska,Pang:2016igs,Deng:2016gyh,Jiang:2016woz},
which leads to the local polarization of hadrons along the vorticity
direction \cite{Becattini:2013fla,Fang:2016vpj} (for a recent review
of the subject, see, e.g., \cite{Wang:2017jpl}).

The global polarization of $\Lambda$ and $\bar{\Lambda}$ has been
measured in the STAR experiment in Au+Au collisions in the collision
energy range 7.7-200 GeV \cite{STAR:2017ckg,Adam:2018ivw} through
their weak decays into pions and protons. The magnitude of the global
polarization is about a few percent and decreases with increasing
collision energies. Hydrodynamic and transport models have been proposed
to describe the polarization data for $\Lambda$ and $\bar{\Lambda}$
from which the vorticity fields can be determined \cite{Baznat:2013zx,Csernai:2013bqa,Csernai:2014ywa,Teryaev:2015gxa,Jiang:2016woz,Deng:2016gyh,Ivanov:2017dff,Li:2017slc,Wei:2018zfb}.
Then, through an integration of the vorticity over the freezeout hyper-surface
\cite{Becattini:2013fla,Fang:2016vpj}, the global polarization of
$\Lambda$ and $\bar{\Lambda}$ is obtained and agrees with the data
\cite{Karpenko:2016jyx,Xie:2017upb,Li:2017slc,Sun:2017xhx,Wei:2018zfb}.

The previous STAR measurement of the global polarization is limited
to the central rapidity region. How the polarization behaves in the
forward rapidity region can shed light on the polarization mechanism.
The STAR collaboration are currently working on a series of upgrades
in the forward region, which will add calorimetry and charged-particle
tracking in the rapidity range $[2.5,4]$, and are expected to collect
the data of Au+Au collisions at 200 GeV in 2023. Then $\Lambda$ and
$\bar{\Lambda}$ may be constructed in this forward region, allowing
for the measurement of their polarization.

In this paper, we will give a geometric model for the hadron polarization
with an emphasis on the rapidity dependence. This work is the natural
extension of a previous work by some of us \cite{Gao:2007bc}. The
geometric model is based on the model of Brodsky, Gunion, and Kuhn
(BGK) \cite{Brodsky:1977de} and that of the Bjorken scaling \cite{Gao:2007bc}.
The BGK model can give a good description of the hadron's rapidity
distribution in nucleus-nucleus collisions.

The paper is organized as follows. In Sect. II, we give formulas for
the average longitudinal momentum and local orbital angular momentum
using the method of Ref. \cite{Gao:2007bc} where the rapidity distribution
of hadrons is given by the BGK model. In Sect. III, we use the hard
sphere and Woods-Saxon model for the nuclear density distribution
to calculate the rapidity distribution of hadrons. In Sect. IV, the
hadron polarization from the local orbital angular momentum is calculated
with the WS nuclear density distribution. By constraining the parameter
by the polarization data at mid-rapidity, we make predictions of the
polarization in the forward rapidity region. The summary is given
in the last section.

\section{Average longitudinal momentum and local orbital angular momentum}

\begin{figure}
\includegraphics[scale=0.8]{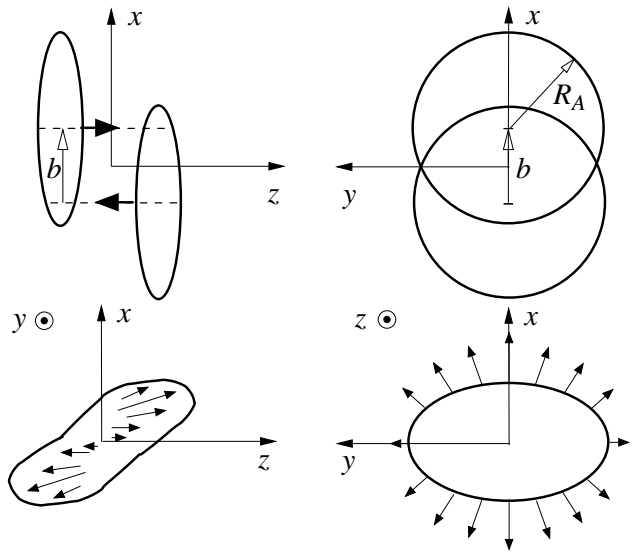}

\caption{The schematic figure taken from \cite{Gao:2007bc} for non-central
heavy ion collisions with impact parameter $\mathbf{b}$ pointing
to $x$ direction. The orbital angular momentum is in $-y$ direction.
\label{fig:no-central-hic}}
\end{figure}

There is an intrinsic rotation of the initially produced matter in
the reaction plane in non-central heavy ion collisions. The rotation
can be characterized by tilted local rapidity distribution of produced
hadrons toward the projectile and target direction in the transverse
plane. We consider non-central collisions of two nuclei $A+A$: the
first one is regarded as the projectile moving in $z$ direction while
the second is regarded as the target moving in $-z$ direction, see
Fig. \ref{fig:no-central-hic} for illustration. The impact parameter
is in the direction from the target to the projectile, i.e. in $x$
direction. The orbital angular momentum (OAM) is in the direction
that is determined by the vector product of the impact parameter and
the projectile momentum, $\mathbf{b}\times\mathbf{p}_{\mathrm{proj}}$
which is $-y$ direction.

In the center of the rapidity frame of $\mathrm{p}+A$ collisions,
the proton has rapidity $Y_{L}$ and interacts at a transverse impact
parameter $\mathbf{r}_{T}$ with $N_{\mathrm{Part}}^{A}\approx\sigma_{NN}T_{A}(\mathbf{r}_{T})$
nucleons with rapidity $-Y_{L}$, where $T_{A}(\mathbf{r}_{T})$ is
the thickness function or the number of nucleons per unit area 
\begin{equation}
T_{A}(\mathbf{r}_{T})=\int dz\rho_{A}(\mathbf{r}),
\end{equation}
with $\mathbf{r}=(x,y,z)$, $\mathbf{r}_{T}=(x,y)$, $\rho_{A}$ is
the number density of nucleons in the nucleus, $\sigma_{NN}$ is the
inelastic cross section of nucleon-nucleon collisions, and $Y_{L}\approx\ln[\sqrt{s}/(2m_{N})]$
is the largest rapidity. The trangular rapidity distribution of hadrons
is the result of string fragmentation between the projectile proton
and the target nucleus. The hadrons produced by the wounded projectile
proton are in the rapidity range $Y\in[0,Y_{L}]$, while those produced
by the wounded target nucleons are in the range $Y\in[-Y_{L},0]$.
The rapidity distribution of produced hadrons is approximately given
by the BGK model \cite{Brodsky:1977de} as 
\begin{eqnarray}
\frac{d^{3}N_{\mathrm{p}A}}{d^{2}\mathbf{r}_{T}dY} & = & \frac{dN_{\mathrm{pp}}}{dY}\left[T_{A}(\mathbf{r}_{T})\frac{Y_{L}-Y}{2Y_{L}}+T_{\mathrm{p}}\frac{Y_{L}+Y}{2Y_{L}}\right],\label{eq:pa-dist}
\end{eqnarray}
where $T_{\mathrm{p}}\approx1$ is the number of projectile protons
per unit area. In the forward or projectile region $Y\approx Y_{L}$
the rapidity distribution approaches that of p+p collisions $dN_{\mathrm{pp}}/dY$,
while in the backward or target region it approaches $(dN_{\mathrm{pp}}/dY)T_{A}(\mathbf{r}_{T})$.
From experimental data we take a Gaussian form for $dN_{\mathrm{pp}}/dY$,
\begin{equation}
\frac{dN_{\mathrm{pp}}}{dY}=a_{1}\,\exp\left(-\frac{Y^{2}}{a_{2}}\right)\frac{1}{\sqrt{1+a_{3}(\cosh Y)^{4}}},\label{eq:pp-y-dist}
\end{equation}
where $a_{1}$, $a_{2}$, $a_{3}$ are parameters. The values of these
parameters in inelastic non-diffractive events in p+p collisions at
some collision energies are determined from the simulation of PYTHIA8.2
\cite{Sjostrand:2014zea} and are listed in Table \ref{tab:parameters-pp}.

\begin{table}[H]
\caption{The values of parameters in the hadron rapidity distribution in inelastic
non-diffractive events in p+p collisions at various collision energies
from the simulation of PYTHIA8.2 \cite{Sjostrand:2014zea}. \label{tab:parameters-pp}}

\begin{centering}
\begin{tabular}{|c|c|c|c|c|c|}
\hline 
$\sqrt{s_{NN}}$ (GeV) & 200 & 130 & 62.4 & 54.4 & 39\tabularnewline
\hline 
$a_{1}$ & 4.584 & 4.096 & 3.862 & 3.726 & 3.420\tabularnewline
\hline 
$a_{2}$ & 26.112 & 25.896 & 18.911 & 18.931 & 18.779\tabularnewline
\hline 
$a_{3}$ & 9.70$\times10^{-8}$ & 5.61$\times10^{-7}$ & 9.75$\times10^{-6}$ & 1.71$\times10^{-5}$ & 6.61$\times10^{-5}$\tabularnewline
\hline 
\end{tabular}
\par\end{centering}

\centering{}%
\begin{tabular}{|c|c|c|c|c|c|}
\hline 
$\sqrt{s_{NN}}$ (GeV) & 27 & 19.6 & 14.5 & 11.5 & 7.7\tabularnewline
\hline 
$a_{1}$ & 3.421 & 3.099 & 3.049 & 2.784 & 2.831\tabularnewline
\hline 
$a_{2}$ & 13.555 & 13.629 & 9.947 & 10.488 & 8.008\tabularnewline
\hline 
$a_{3}$ & 2.50$\times10^{-4}$ & 8.76$\times10^{-4}$ & 2.44$\times10^{-3}$ & 5.90$\times10^{-3}$ & 9.40$\times10^{-3}$\tabularnewline
\hline 
\end{tabular}
\end{table}

Such a trapezoidal shape of the rapidity distribution in (\ref{eq:pa-dist})
in the BGK model is a consequence of the string fragmentation and
can be described naturally by the LUND string \cite{Andersson:1986gw}
and HIJING model \cite{Wang:1991hta,Gyulassy:1994ew}. An extension
of the BGK model has been applied to the jet tomography of twisted
strongly-coupled quark-qluon plasmas \cite{Adil:2005qn} as well as
the global polarization in nucleus-nucleus collisions \cite{Betz:2007kg}.
In nucleus-nucleus collisions with projectile $A$ and target $B$,
at the point $\mathbf{r}_{T}=(x,y)$ in the transverse plane in the
participant region (the coordinate system is shown in the upper-left
of Fig. \ref{fig:no-central-hic}), the rapidity distribution of produced
hadrons has the form which is a generalization of Eq. (\ref{eq:pa-dist}),
i.e. the sum over contributions from projectile ('proj') and target
('tar') 
\begin{eqnarray}
\frac{d^{3}N_{AB}}{d^{2}\mathbf{r}_{T}dY} & = & \frac{d^{3}N_{A}^{\mathrm{proj}}}{d^{2}\mathbf{r}_{T}dY}+\frac{d^{3}N_{B}^{\mathrm{tar}}}{d^{2}\mathbf{r}_{T}dY}\nonumber \\
 & = & \frac{dN_{\mathrm{pp}}}{dY}\left[T_{A}(\mathbf{r}_{T}-\mathbf{b}/2)\frac{Y_{L}+Y}{2Y_{L}}+T_{B}(\mathbf{r}_{T}+\mathbf{b}/2)\frac{Y_{L}-Y}{2Y_{L}}\right].\label{eq:aa-dist}
\end{eqnarray}
Here, thickness functions $T_{A}(\mathbf{r}_{T}-\mathbf{b}/2)$ and
$T_{B}(\mathbf{r}_{T}+\mathbf{b}/2)$ in Eq. (\ref{eq:aa-dist}) are
given by 
\begin{eqnarray}
T_{A,B}(\mathbf{r}_{T}\mp\mathbf{b}/2) & = & \int dz\rho^{A,B}(\mathbf{r}_{T}\mp\mathbf{b}/2),\label{eq:thickness-func}
\end{eqnarray}
where $\rho^{A,B}(\mathbf{r}_{T}\mp\mathbf{b}/2)$ the participant
nucleon number density functions of nuclei $A$ and $B$. One can
check that the distribution (\ref{eq:aa-dist}) is proportional to
$T_{A/B}(\mathbf{r}_{T}\mp\mathbf{b}/2)$ at the $Y=\pm Y_{L}$.

From Eq. (\ref{eq:aa-dist}) we can derive the distribution in the
in-plane position $x$ and the rapidity $Y$ by integrating over the
out-plane position $y$ in the range $[-y_{m},y_{m}]$, 
\begin{eqnarray}
\frac{d^{2}N_{AB}}{dxdY} & = & \frac{dN_{\mathrm{pp}}}{dY}\left\{ \frac{1}{2}\int_{-y_{m}}^{y_{m}}dy\left[T_{A}(\mathbf{r}_{T}-\mathbf{b}/2)+T_{B}(\mathbf{r}_{T}+\mathbf{b}/2)\right]\right.\nonumber \\
 &  & \left.+\frac{Y}{2Y_{L}}\int_{-y_{m}}^{y_{m}}dy\left[T_{A}(\mathbf{r}_{T}-\mathbf{b}/2)-T_{B}(\mathbf{r}_{T}+\mathbf{b}/2)\right]\right\} ,\label{eq:aa-dist-1}
\end{eqnarray}
where $y_{m}$ is the maximum of $y$ at a specific $x$, in the hard
sphere model of the nuclear density distribution it is defined by
the boundary of the overlapping region of two nuclei, while in the
Woods-Saxon model there is no sharp boundary but it can be set to
a value much larger than $y_{m}$ in the hard sphere model. We define
the normalized probability distribution of $Y$ at $x$, 
\begin{eqnarray}
f(Y,x) & = & \left(\frac{dN_{AB}}{dx}\right)^{-1}\frac{d^{2}N_{AB}}{dxdY},\label{eq:fyx}
\end{eqnarray}
where the distribution $dN_{AB}/dx$ is given by 
\begin{eqnarray}
\frac{dN_{AB}}{dx} & = & \int_{-Y_{L}}^{Y_{L}}dY\frac{d^{2}N_{AB}}{dxdY}\nonumber \\
 & = & \int_{0}^{Y_{L}}dY\frac{dN_{\mathrm{pp}}}{dY}\int_{-y_{m}}^{y_{m}}dy\left[T_{B}(\mathbf{r}_{T}+\mathbf{b}/2)+T_{A}(\mathbf{r}_{T}-\mathbf{b}/2)\right].\label{eq:dNABdx}
\end{eqnarray}

According to the Bjorken scaling model \cite{Gao:2007bc}, the average
rapidity of the particle as a function of $Y$ at $x$ in the rapidity
window $[Y-\Delta_{Y}/2,Y+\Delta_{Y}/2]$ is given by 
\begin{eqnarray}
\left\langle Y\right\rangle _{\Delta} & = & \frac{\int_{Y-\Delta_{Y}/2}^{Y+\Delta_{Y}/2}dY^{\prime}\:Y^{\prime}f(Y^{\prime},x)}{\int_{Y-\Delta_{Y}/2}^{Y+\Delta_{Y}/2}dY^{\prime}\:f(Y^{\prime},x)}\nonumber \\
 & \approx & Y+\frac{\Delta_{Y}^{2}}{12}\frac{1}{f(Y,x)}\frac{df(Y,x)}{dY},\label{eq:ycx}
\end{eqnarray}
where $\Delta_{Y}$ is the width of the rapidity window in which particles
interact to reach collectivity. We assumed $\Delta_{Y}\ll Y$ so $\Delta_{Y}$
can be treated as a perturbation. The average rapidity of the particle
as a function of $x$ in the full rapidity range reads 
\begin{eqnarray}
\left\langle Y\right\rangle  & = & \frac{\int_{-Y_{L}}^{Y_{L}}dY\:Yf(Y,x)}{\int_{-Y_{L}}^{Y_{L}}dY\:f(Y,x)}\nonumber \\
 & = & \frac{1}{Y_{L}}\left\langle Y^{2}\right\rangle _{\mathrm{pp}}\frac{\int_{-y_{m}}^{y_{m}}dy\left[T_{A}(\mathbf{r}_{T}-\mathbf{b}/2)-T_{B}(\mathbf{r}_{T}+\mathbf{b}/2)\right]}{\int_{-y_{m}}^{y_{m}}dy\left[T_{A}(\mathbf{r}_{T}-\mathbf{b}/2)+T_{B}(\mathbf{r}_{T}+\mathbf{b}/2)\right]},\label{eq:ycx-1}
\end{eqnarray}
where $\left\langle Y^{2}\right\rangle _{\mathrm{pp}}$ is defined
as 
\begin{equation}
\left\langle Y^{2}\right\rangle _{\mathrm{pp}}=\frac{\int_{-Y_{L}}^{Y_{L}}dY\:(dN_{\mathrm{pp}}/dY)Y^{2}}{\int_{-Y_{L}}^{Y_{L}}dY\:(dN_{\mathrm{pp}}/dY)}.
\end{equation}
The average longitudinal momentum $p_{z}$ and the average energy
$E_{p}$ of the particle are 
\begin{eqnarray}
\left\langle p_{z}\right\rangle  & = & \left\langle p_{T}\right\rangle \sinh\left\langle Y\right\rangle _{\Delta}\nonumber \\
 & \approx & \left\langle p_{T}\right\rangle \sinh Y+\left\langle p_{T}\right\rangle \frac{\Delta_{Y}^{2}}{12}\frac{d\ln f(Y,x)}{dY}\cosh Y,\nonumber \\
\left\langle E_{p}\right\rangle  & = & \left\langle p_{T}\right\rangle \cosh\left\langle Y\right\rangle _{\Delta}\nonumber \\
 & \approx & \left\langle p_{T}\right\rangle \cosh Y+\left\langle p_{T}\right\rangle \frac{\Delta_{Y}^{2}}{12}\frac{d\ln f(Y,x)}{dY}\sinh Y,\label{eq:pz-Y}
\end{eqnarray}
where we have treated terms proportional to $\Delta_{Y}^{2}$ as a
perturbation. At a given $Y$ we consider two particles located at
$x+\Delta x/2$ and $x-\Delta x/2$, in their center of mass frame,
the local average OAM for two colliding particles is given by \cite{Gao:2007bc}
\begin{eqnarray}
\left\langle L_{y}\right\rangle  & \approx & -(\Delta x)\left\langle p_{z}^{\mathrm{cm}}\right\rangle \nonumber \\
 & \approx & -(\Delta x)^{2}\left\langle p_{T}\right\rangle \frac{\Delta_{Y}^{2}}{24}\frac{d\ln f(Y,x)}{dYdx},\label{eq:Ly}
\end{eqnarray}
where $\left\langle p_{z}^{\mathrm{cm}}\right\rangle $ is the average
longitudinal momentum in the center of mass frame for one particle.
Here $\Delta x$ is a typical impact parameter of particle scatterings.
In following sections we will use the average of $d\ln f(Y,x)/dxdY$
over the in-plane coordinate 
\begin{equation}
\left\langle \frac{d\ln f(Y,x)}{dYdx}\right\rangle =\frac{\int dx(dN_{AB}/dxdY)(d\ln f(Y,x)/dYdx)}{\int dx(dN_{AB}/dxdY)},\label{eq:dlnfdxdY_ave}
\end{equation}
where $dN_{AB}/dxdY$ is given in Eq. (\ref{eq:aa-dist-1}) as a weight
function.

\section{Rapidity distributions of hadrons in hard sphere and Woods-Saxon
model}

In this section we will calculate the rapidity distributions for hadrons
$f(Y,x)$ in Eq. (\ref{eq:fyx}) with the hard sphere (HS) and Woods-Saxon
(WS) nuclear density distribution, which are involved in the thickness
functions in Eq. (\ref{eq:thickness-func}). As a simple illustration,
we consider collisions of two identical nuclei with nucleon number
$A$.

\subsection{Hard sphere nuclear density distribution}

\begin{figure}[H]
\centering\includegraphics[scale=0.4]{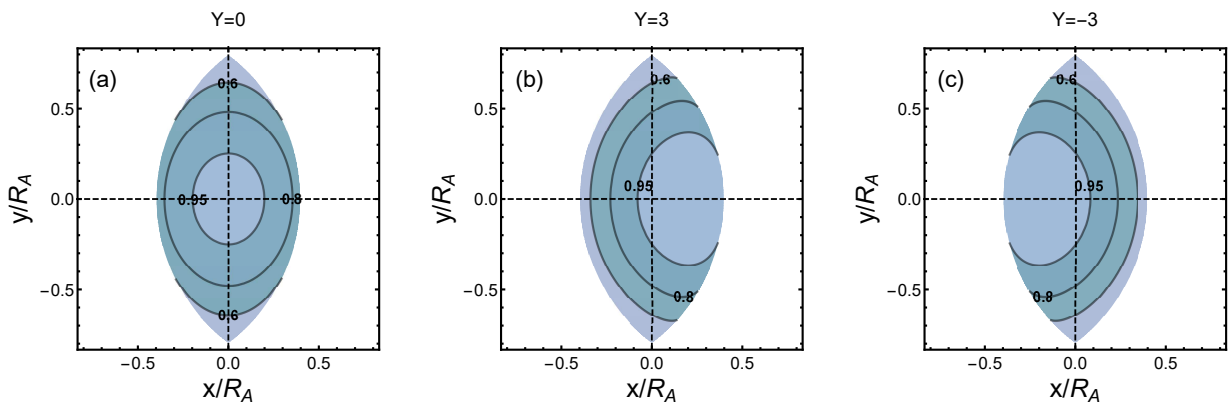}\caption{The hadron distributions (contour plot) in the HS model in the transverse
plane for Au+Au collisions at $\sqrt{s_{NN}}=200$ GeV and $b=1.2R_{A}$.
The number on the contour line denotes the value on the line normalized
by that at the origin. The rapidity values are chosen to be $Y=0$
(central), $Y=3$ (forward) and $Y=-3$ (backward). \label{fig:dNaadxdy_hs}}
\end{figure}

The HS nuclear density is given by 
\begin{equation}
\rho_{\mathrm{HS}}(\mathbf{r})=\frac{3A}{4\pi R_{A}^{3}}\theta(R_{A}-r),
\end{equation}
where $R_{A}=1.2A^{1/3}$ fm is the nucleus radius. The thickness
functions have the analytical form 
\begin{eqnarray}
T_{A}(\mathbf{r}_{T}\pm\mathbf{b}/2) & = & \frac{6A}{4\pi R_{A}^{3}}\left[R_{A}^{2}-(x\pm b/2)^{2}-y^{2}\right]^{1/2}.
\end{eqnarray}
Inserting the above into Eq. (\ref{eq:aa-dist}), we obtain the hadron
distribution $dN_{AA}/(dxdydY)$, whose numerical results are shown
in Fig. \ref{fig:dNaadxdy_hs} at three rapidity values in Au+Au collisions
at $\sqrt{s_{NN}}=200$ GeV with the impact parameter $b=1.2R_{A}$.
In the HS model, the overlapping region of two nuclei is limited by
$|x|<R_{A}-b/2$ and $|y|<\sqrt{R_{A}^{2}-(|x|+b/2)^{2}}$. We see
that the distribution at $Y=0$ is symmetric in $x$ and $y$ while
that the distribution in the forward (backward) rapidity is shifted
to the right (left) in $x$ direction.

\begin{figure}[H]
\centering\includegraphics[scale=0.4]{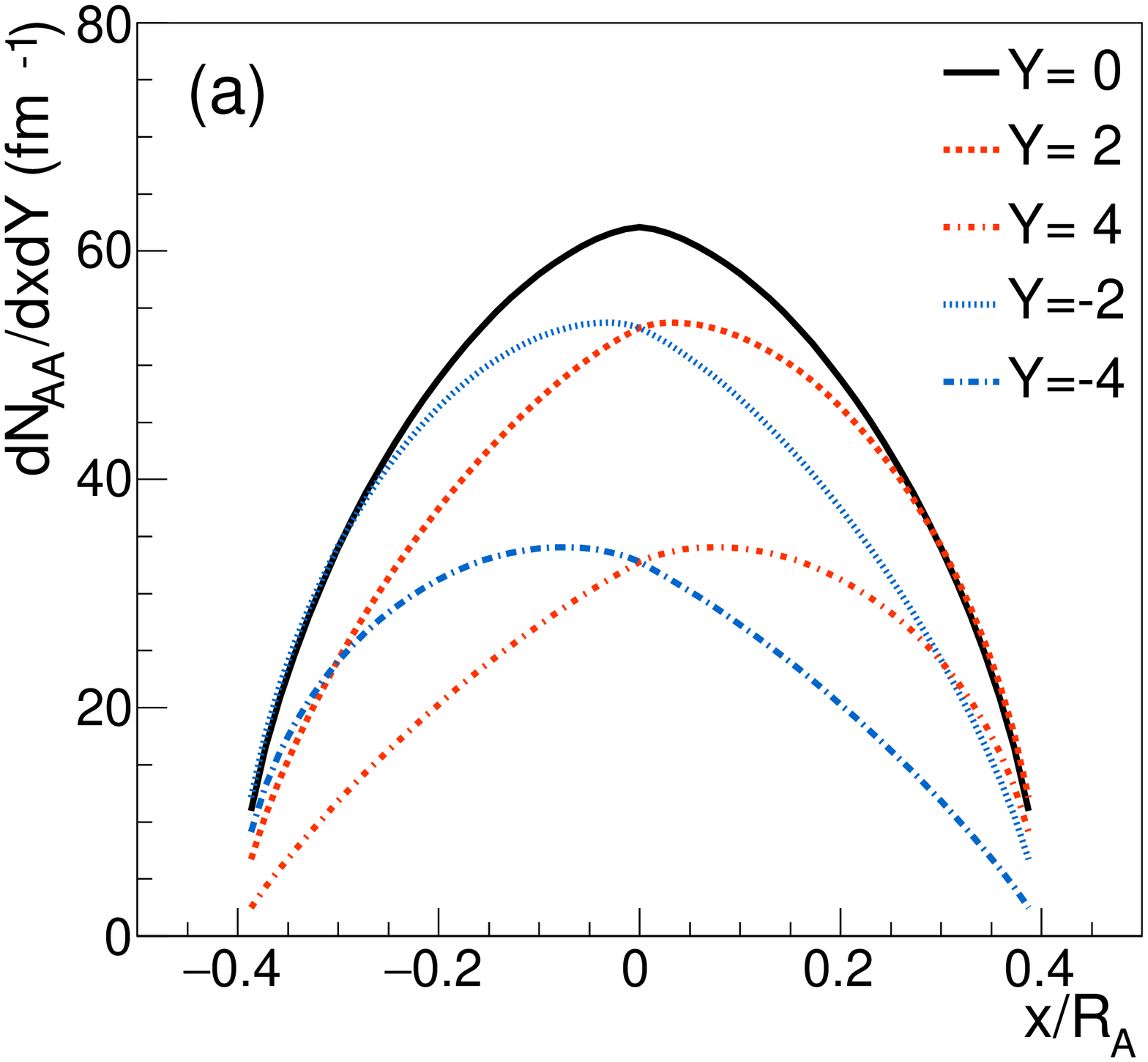}\includegraphics[scale=0.4]{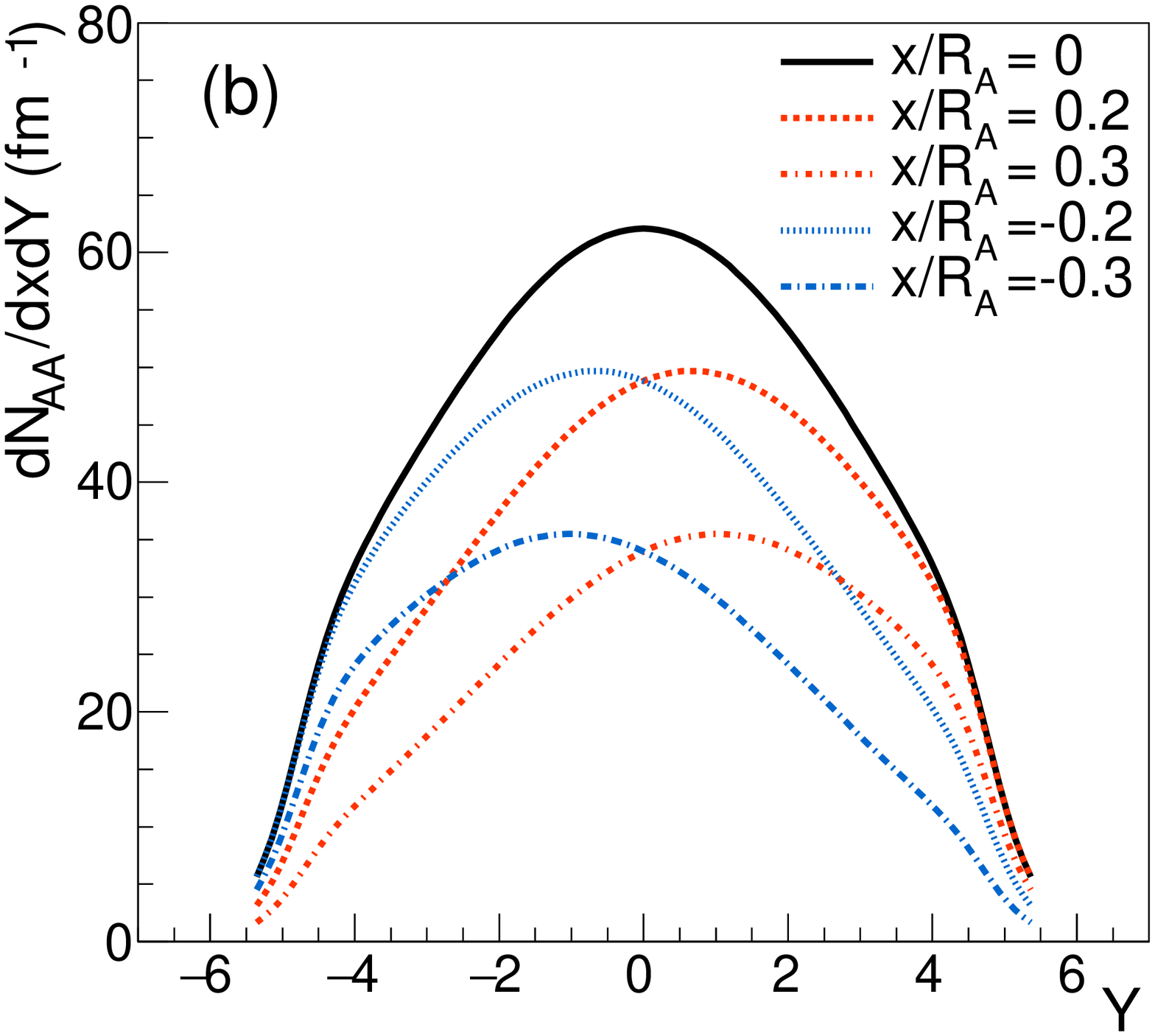}

\centering\includegraphics[scale=0.4]{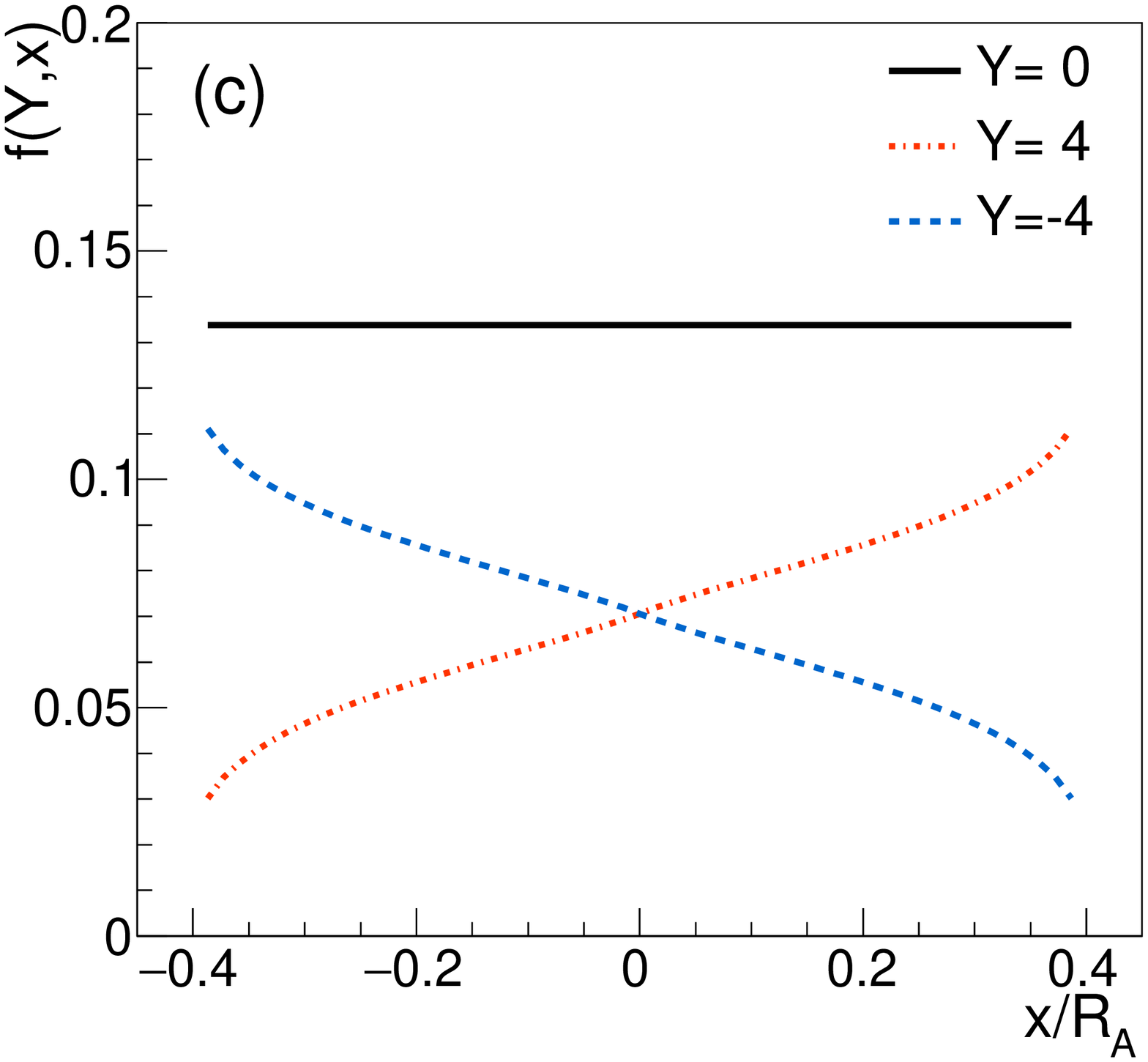}\includegraphics[scale=0.4]{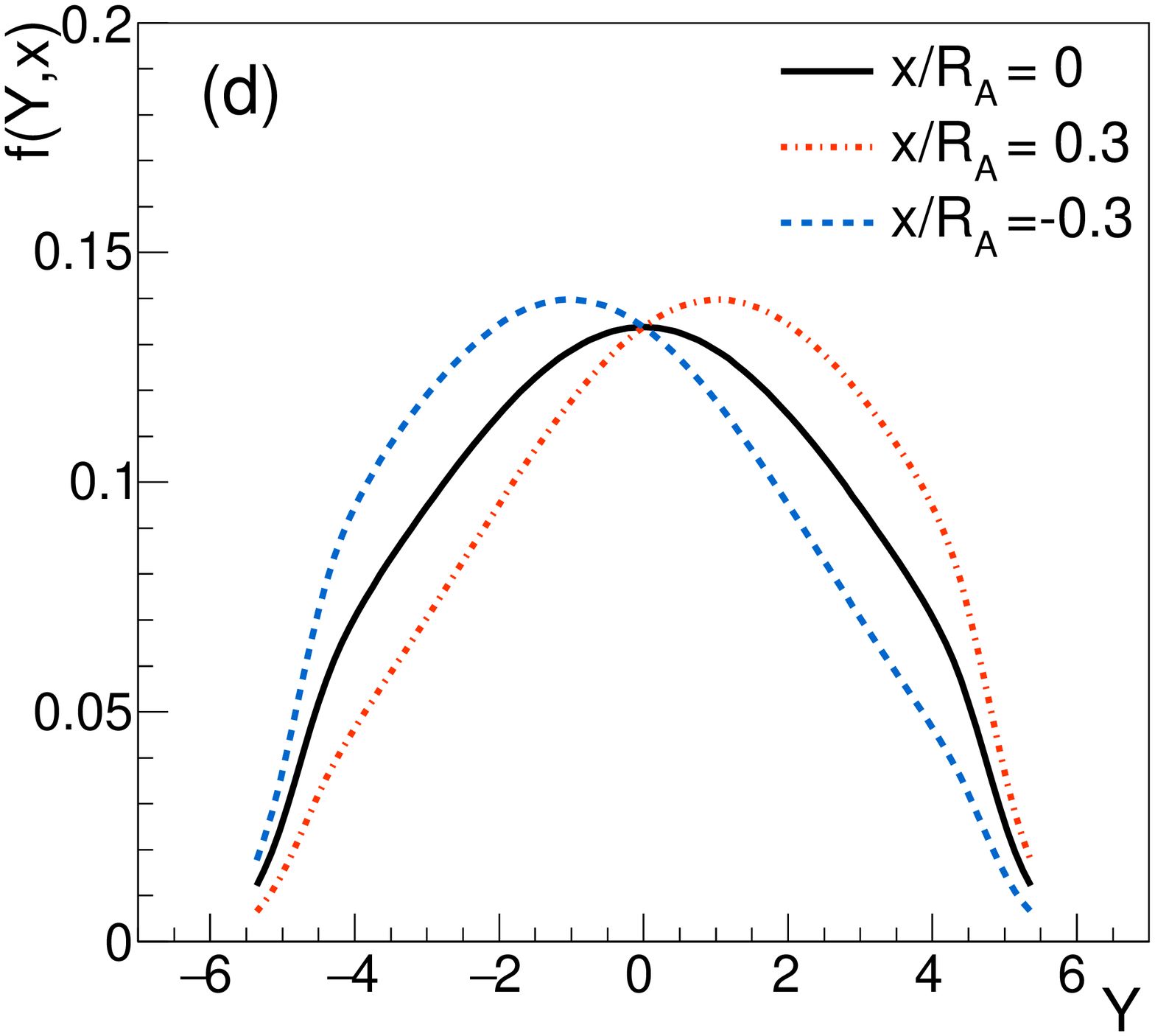}

\caption{The hadron distributions in the HS model in Au+Au collisions at $\sqrt{s_{NN}}=200$
GeV as functions of (a) the in-plane position $x$ at different rapidity
values and as functions of (b) the rapidity $Y$ at different values
of $x$. The impact parameter is set to $b=1.2R_{A}$. (c) The normalized
distribution $f(Y,x)$ as functions of $x$ at different $Y$. (d)
The normalized distribution $f(Y,x)$ corresponding to (b). The definition
of $f(Y,x)$ is given in Eq. (\ref{eq:fyx-hs}). \label{fig:dNaadx_hs}}
\end{figure}

Integrating over the out-plane coordinate $y$, we obtain the hadron
distribution function 
\begin{equation}
\frac{d^{2}N_{AA}}{dxdY}=\frac{6A}{4\pi R_{A}^{3}}\frac{dN_{pp}}{dY}\left[C_{1}^{+}+C_{1}^{-}+\frac{Y}{Y_{L}}\left(C_{1}^{+}-C_{1}^{-}\right)\right],\label{eq:dnaa-dxdY}
\end{equation}
where $C_{1}^{\pm}$ are defined in Eq. (\ref{eq:c1_define}). In
Fig. \ref{fig:dNaadx_hs}(a), we show $dN_{AA}/dxdY$ as functions
of in-plane coordinate $x$ at various rapidity values. We see that
the distribution at $Y=0$ is symmetric while the distribution at
forward (backward) rapidity is shifted to the positive (negative)
$x$. The magnitude of the shift increases slightly with the rapidity.
In Fig. \ref{fig:dNaadx_hs}(b), we show $dN_{AA}/dxdY$ as functions
of the rapidity at different $x$. We see that the distribution at
$x=0$ is symmetric while that at positive (negative) $x$ is tilted
to the forward (backward) rapidity. From Eq. (\ref{eq:fyx}), we obtain
the normalized function $f(Y,x)$ as 
\begin{eqnarray}
f(Y,x) & = & \frac{dN_{\mathrm{pp}}/dY}{2\int_{0}^{Y_{L}}dY(dN_{\mathrm{pp}}/dY)}\left(1+\frac{Y}{Y_{L}}\cdot\frac{C_{1}^{+}-C_{1}^{-}}{C_{1}^{+}+C_{1}^{-}}\right),\label{eq:fyx-hs}
\end{eqnarray}
where $|x|\leq R_{A}-b/2$ and $b\leq2R_{A}$. The numeical results
of $f(Y,x)$ are shown in Fig. \ref{fig:dNaadx_hs}(c,d). 

\begin{figure}[H]
\centering\includegraphics[scale=0.5]{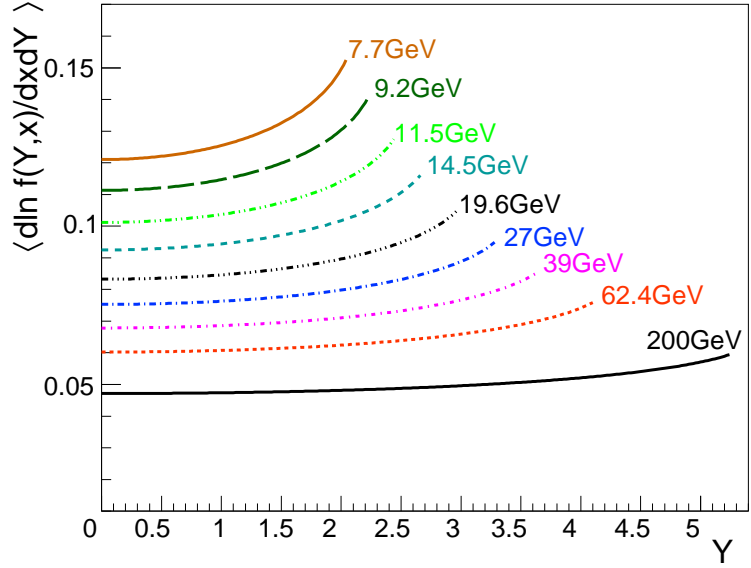}\caption{The average quantity $\left\langle d\ln f(Y,x)/dYdx\right\rangle $
as functions of the rapidity $Y$ in the HS model for Au+Au collisions
at various collision energies. The impact parameter is set to $b=1.2R_{A}$.
\label{fig:dlnfdxdY_ave_hs}}
\end{figure}

The derivative of $\ln f(Y,x)$ with respect to $Y$ and $x$ is derived
in Eq. (\ref{eq:dlnfdydx}), from which one can obtain $\left\langle d\ln f(Y,x)/dxdY\right\rangle $
through Eq. (\ref{eq:dlnfdxdY_ave}). We show the numerical results
of $\left\langle d\ln f(Y,x)/dxdY\right\rangle $ in Fig. \ref{fig:dlnfdxdY_ave_hs}
as rapidity functions in Au+Au collisions at various collision energies.
We see that $\left\langle d\ln f(Y,x)/dYdx\right\rangle $ increases
slowly with the rapidity via the $Y/Y_{L}$ term which can be seen
in Eq. (\ref{eq:dlnfdydx}). The relatively obvious increase in the
forward rapidity region is an artifact of the HS model in comparison
with the WS model in the next subsection. The energy dependence of
$\left\langle d\ln f(Y,x)/dYdx\right\rangle $ is mainly controlled
by $Y_{L}\approx\ln[\sqrt{s}/(2m_{N})]$ as shown in Eq. (\ref{eq:dlnfdydx}). 

\begin{figure}
\centering\includegraphics[scale=0.5]{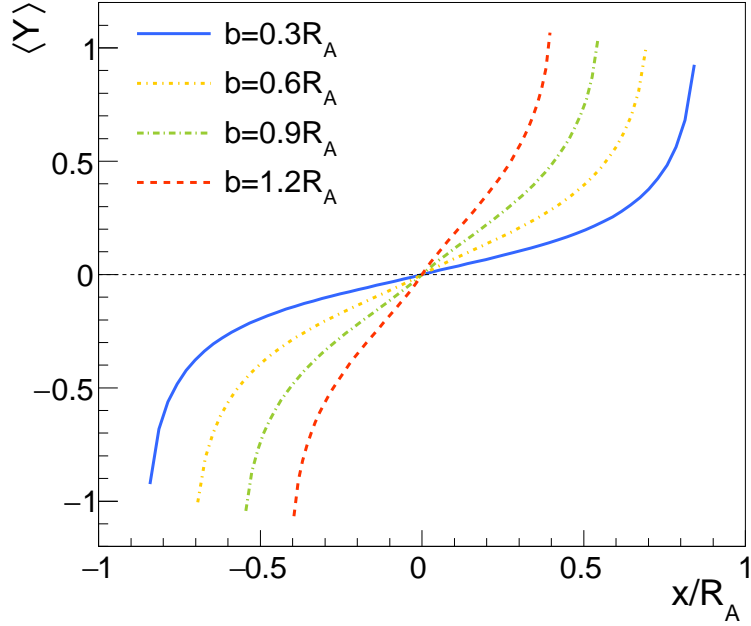}

\caption{The average rapidity in the full rapidity range for Au+Au collisions
at $\sqrt{s_{NN}}=200$ GeV as functions of $x$ from Eq. (\ref{eq:ycx-2})
in the HS model. \label{fig:avY}}
\end{figure}

From Eq. (\ref{eq:ycx-1}) we obtain the average rapidity in the full
rapidity range as 
\begin{eqnarray}
\left\langle Y\right\rangle  & = & \frac{1}{Y_{L}}\left\langle Y^{2}\right\rangle _{\mathrm{pp}}\frac{C_{1}^{+}-C_{1}^{-}}{C_{1}^{+}+C_{1}^{-}}.\label{eq:ycx-2}
\end{eqnarray}
The numerical result of the above average rapidity is shown in Fig.
\ref{fig:avY} which is consistent with Fig. 5 of Ref. \cite{Gao:2007bc}. 

\subsection{Woods-Saxon nuclear density distribution}

In this subsection we choose a more realistic nuclear density distribution,
the WS distribution, defined as 
\begin{equation}
f_{\mathrm{WS}}(\mathbf{r})=\frac{C_{0}}{\exp\left[\left(r-R_{A}\right)/a\right]+1},\label{eq:ws-dist}
\end{equation}
where $r=|\mathbf{r}|$, $a=0.54$ fm, and $C_{0}$ is a normalization
constant to make the volume integral of $f_{\mathrm{WS}}(\mathbf{r})$
to be the number of nucleons in the nucleus, 
\begin{equation}
C_{0}=\frac{A}{4\pi}\left[\int_{0}^{\infty}drr^{2}\frac{1}{\exp[(r-R_{A})/a]+1}\right]^{-1}.
\end{equation}
For Au-197 nuclei, we have $R_{A}\approx6.98$ fm, $C_{0}\approx A/(4\pi)/120.2\approx0.131\:\mathrm{fm}^{-3}$.
According to the Glauber model, the participant nucleon number density
for two colliding nuclei are given by 
\begin{eqnarray}
\rho_{\mathrm{WS}}^{A,B}(\mathbf{r}_{T}\mp\mathbf{b}/2) & = & f_{\mathrm{WS}}^{A,B}(\mathbf{r}_{T}\mp\mathbf{b}/2)\nonumber \\
 &  & \times\left\{ 1-\exp\left[-\sigma_{NN}\int dzf_{\mathrm{WS}}^{B,A}(\mathbf{r}_{T}\pm\mathbf{b}/2)\right]\right\} ,\label{eq:rho_AB_ws}
\end{eqnarray}
where $\sigma_{NN}$ can be taken as the inelastic pp collision cross
section.

\begin{figure}[H]
\centering\includegraphics[scale=0.4]{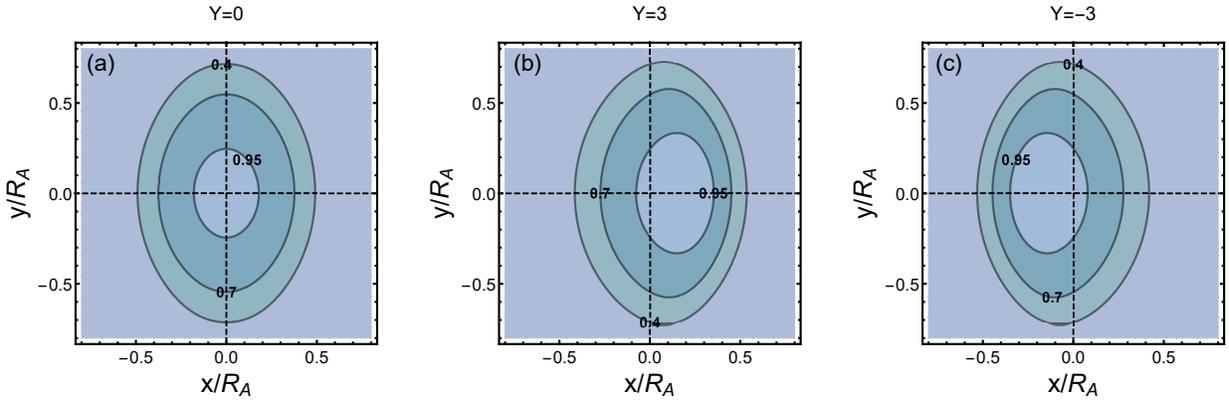}\caption{The hadron distributions (contour plot) in the WS model in the transverse
plane for Au+Au collisions at $\sqrt{s_{NN}}=200$ GeV and $b=1.2R_{A}$.
The number on the contour line denotes the value on the line normalized
by that at the origin. The rapidity values are chosen to be $Y=0$
(central), $Y=3$ (forward) and $Y=-3$ (backward). \label{fig:dNabdxdy_ws}}
\end{figure}

\begin{figure}[H]
\centering\includegraphics[scale=0.37]{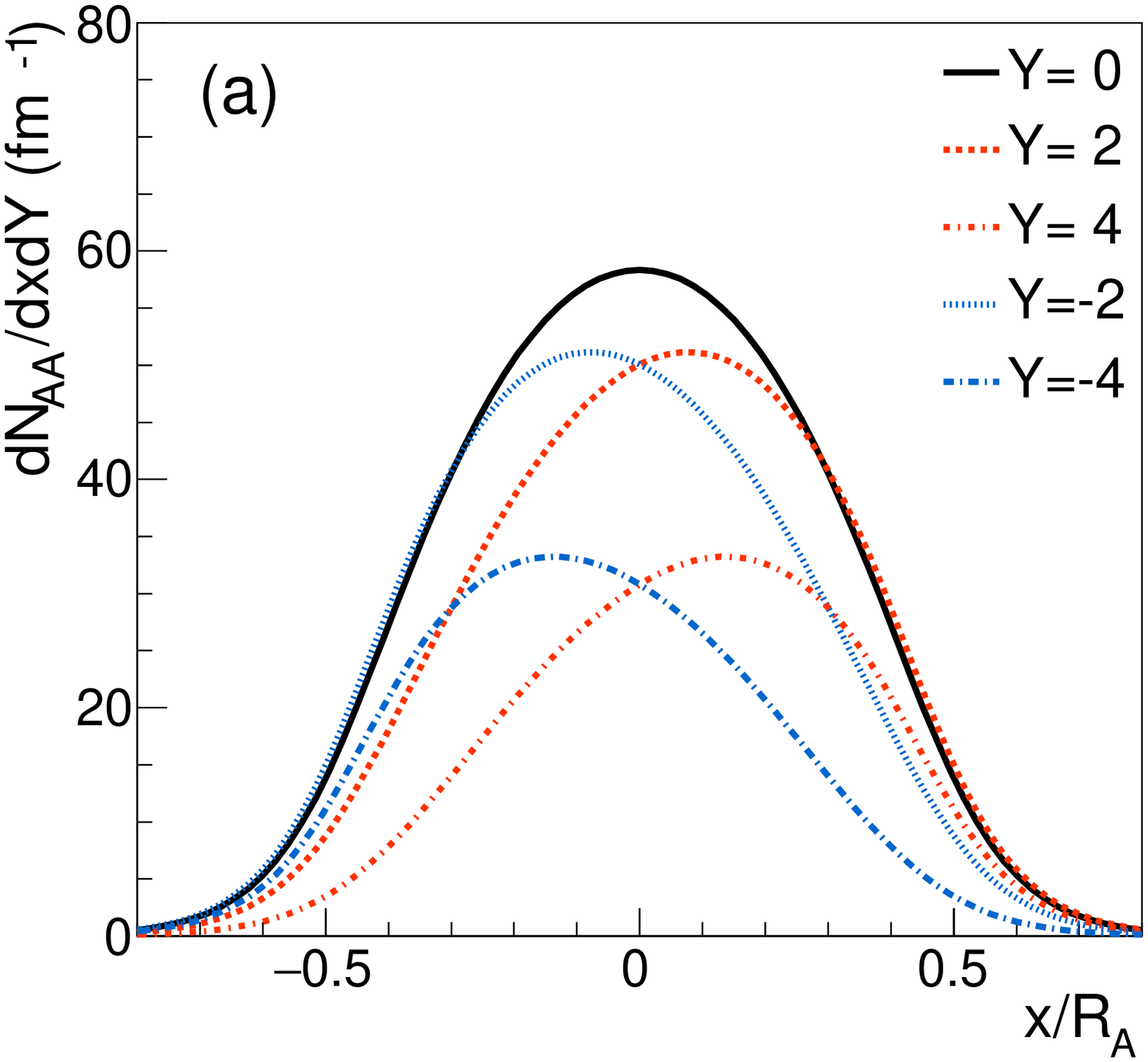}\includegraphics[scale=0.37]{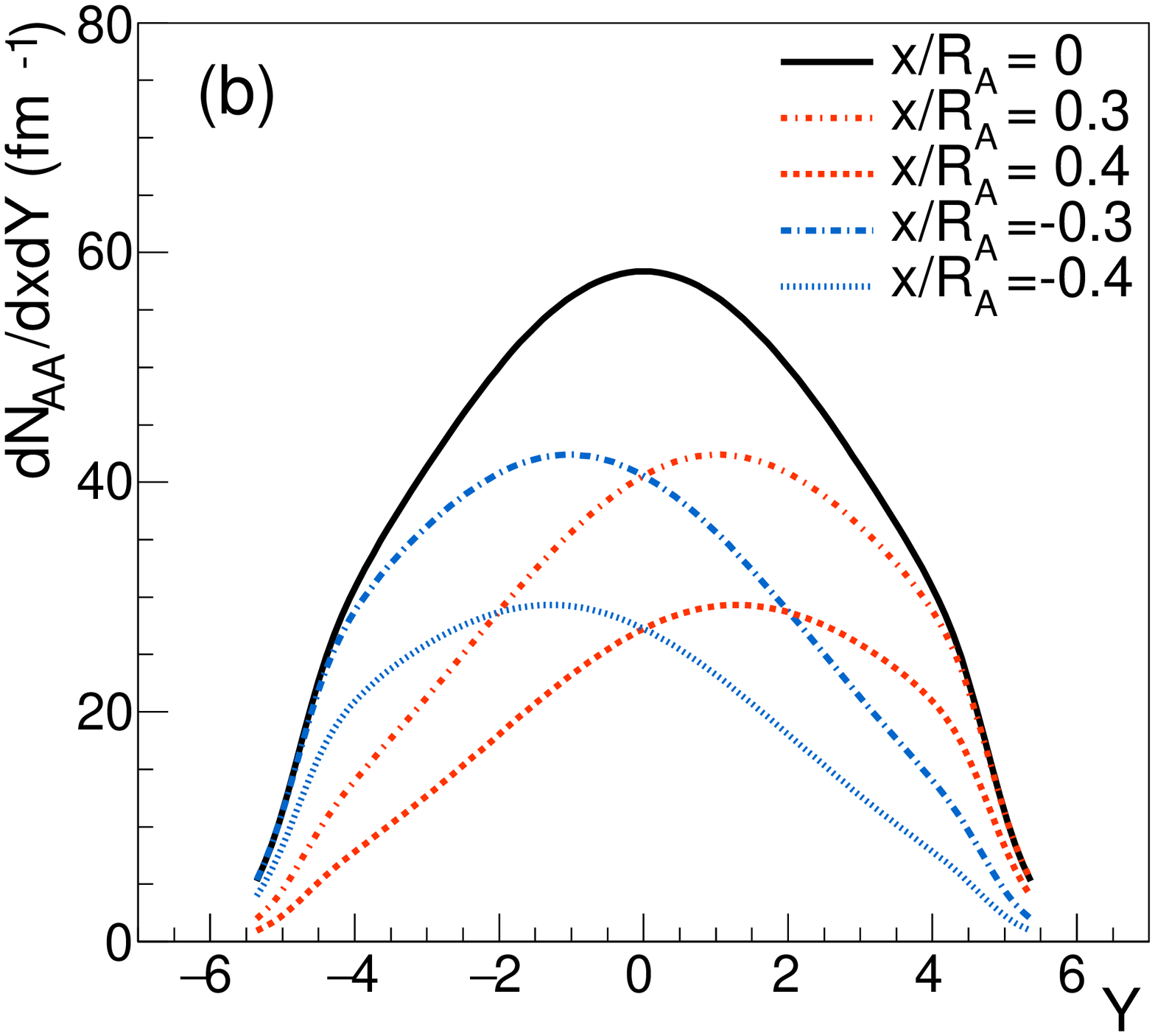}

\centering\includegraphics[scale=0.37]{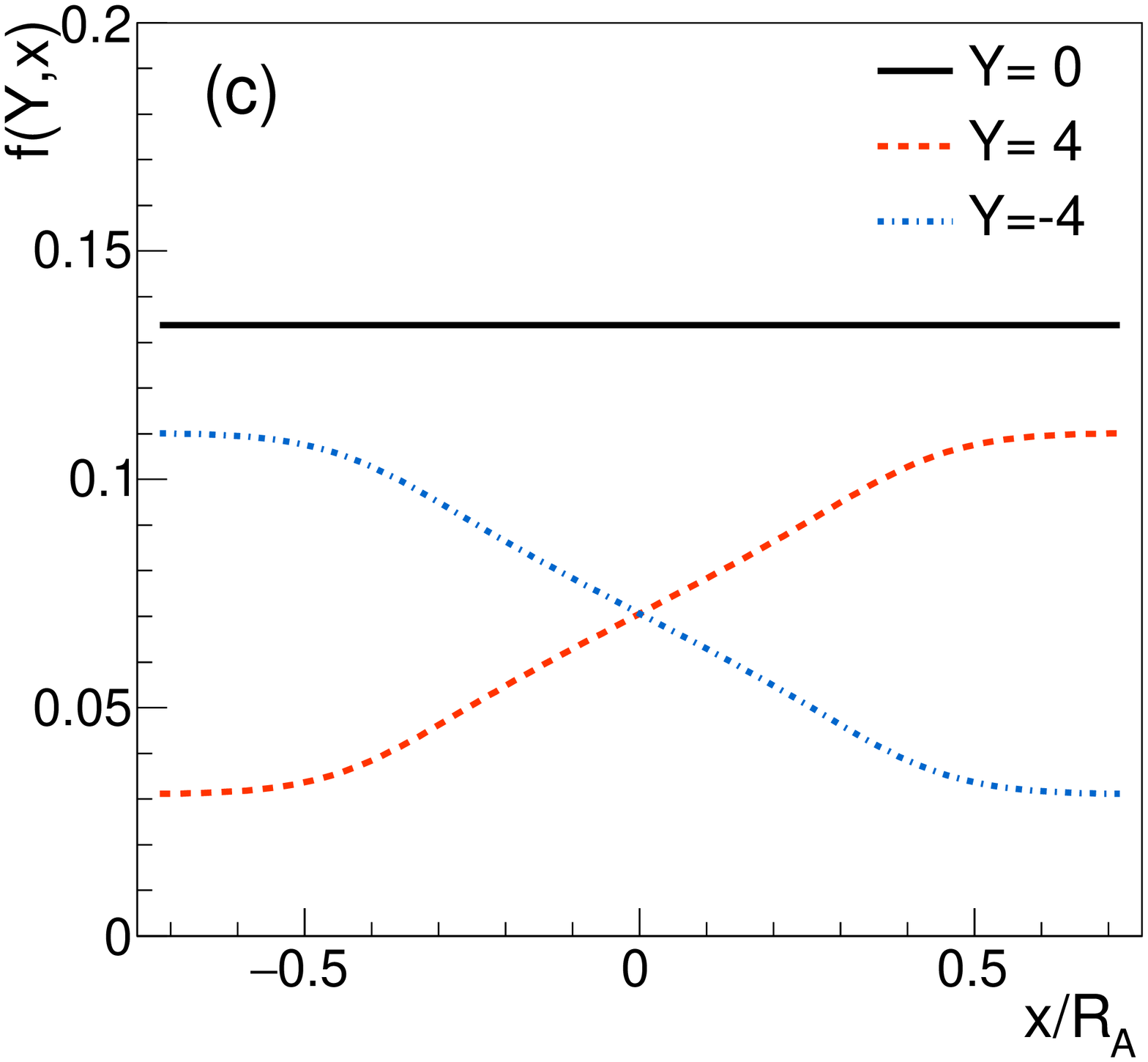}\includegraphics[scale=0.37]{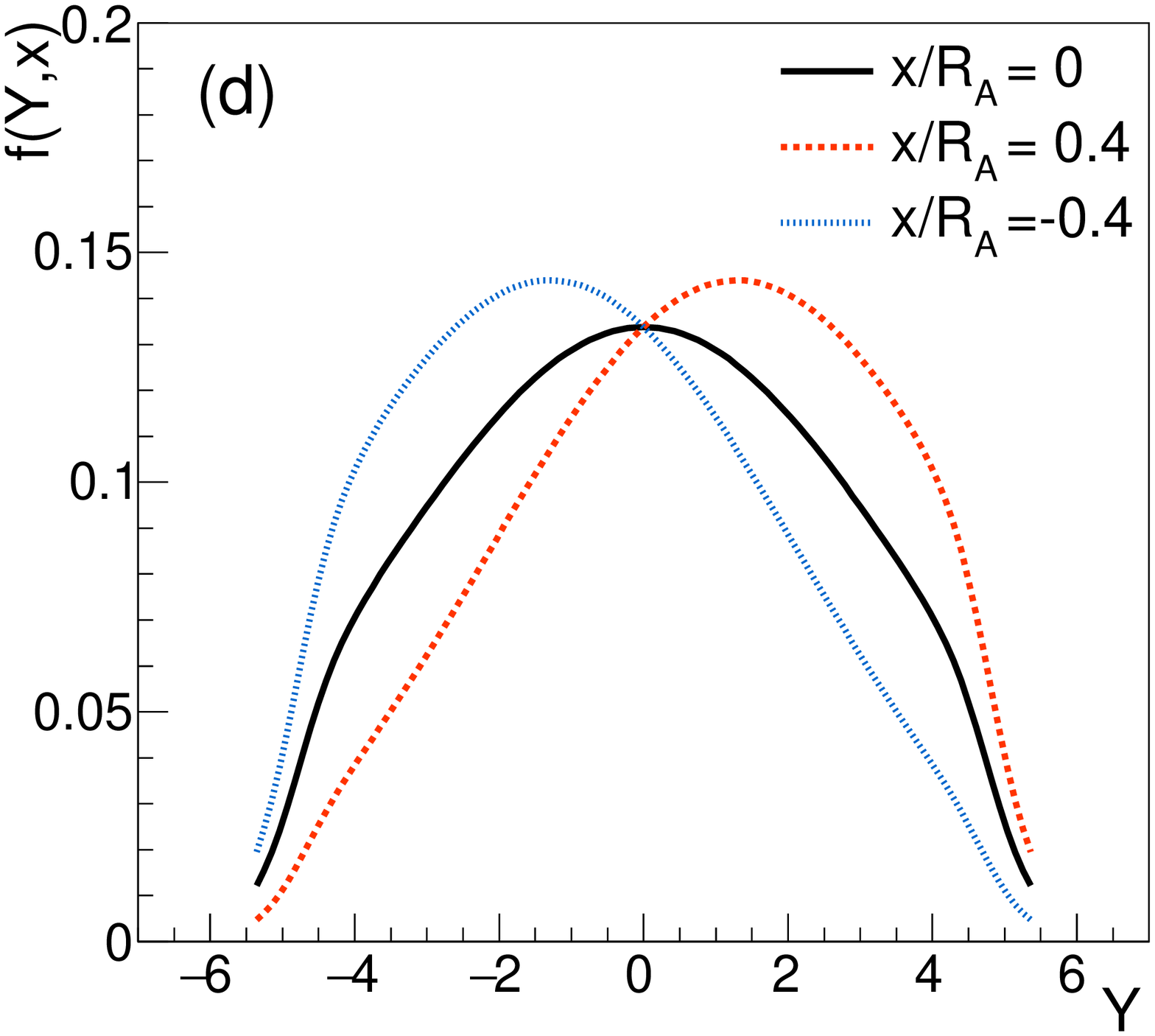}\caption{The hadron distributions in the WS model in Au+Au collisions at $\sqrt{s_{NN}}=200$
GeV as functions of (a) the in-plane position $x$ at different rapidity
values and as functions of (b) the rapidity $Y$ at different values
of $x$. The impact parameter is set to $b=1.2R_{A}$. (c) The normalized
distribution $f(Y,x)$ as functions of $x$ at different $Y$. (d)
The normalized distribution $f(Y,x)$ corresponding to (b). The definition
of $f(Y,x)$ is given in Eq. (\ref{eq:fyx}). \label{fig:dNaadxdY_ws}}
\end{figure}

We consider collisions of two identical nuclei. With the WS distribution
in Eq. (\ref{eq:ws-dist}), we can calculate $f(Y,x)$ in Eq. (\ref{eq:fyx}).
From Eq. (\ref{eq:thickness-func}), the thickness functions become
\begin{eqnarray}
T_{A}(\mathbf{r}_{T}\pm\mathbf{b}/2) & = & \int_{-\infty}^{\infty}dz\rho_{\mathrm{WS}}^{A}(\mathbf{r}_{T}\pm\mathbf{b}/2)\nonumber \\
 & = & \int_{-\infty}^{\infty}dzf_{\mathrm{WS}}(\mathbf{r}_{T}\pm\mathbf{b}/2)\nonumber \\
 &  & \times\left\{ 1-\exp\left[-\sigma_{NN}\int dzf_{\mathrm{WS}}(\mathbf{r}_{T}\mp\mathbf{b}/2)\right]\right\} .\label{eq:TA_ws}
\end{eqnarray}
Substituting Eq. (\ref{eq:TA_ws}) into Eq. (\ref{eq:aa-dist}) and
(\ref{eq:aa-dist-1}), we obtain the hadron distribution $dN_{AA}/(dxdydY)$
in $xy$ plane and $dN_{AA}/(dxdY)$ by an integration over $y$,
respectively. The numerical results for $dN_{AA}/(dxdydY)$ and $dN_{AA}/(dxdY)$
are shown in Fig. \ref{fig:dNabdxdy_ws} and \ref{fig:dNaadxdY_ws}
respectively for different rapidity values in Au+Au collisions at
200 GeV with $b=1.2R_{A}$. Similar to results of the HS model in
Fig. \ref{fig:dNaadxdy_hs} and (\ref{fig:dNaadx_hs}), in the forward/backward
rapidity region, the hadron distributions are tilted toward the positive/negative
$x$. But different from results of the HS model, the hadron distributions
in the WS model are smooth in the edge of the overlapping region.

\begin{figure}[H]
\centering\includegraphics[scale=0.5]{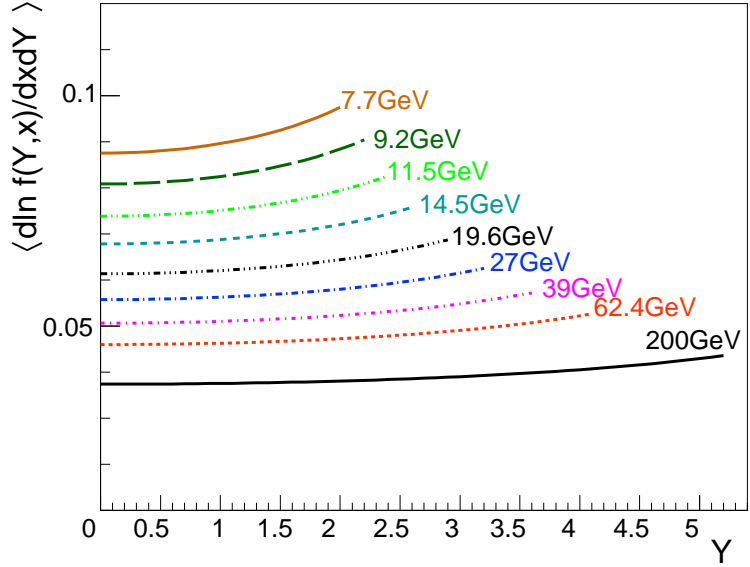}\caption{The average quantity $\left\langle d\ln f(Y,x)/dYdx\right\rangle $
as functions of the rapidity $Y$ in the WS model for Au+Au collisions
at various collision energies. The impact parameter is set to $b=1.2R_{A}$.
\label{fig:dlnfdxdY_ave_ws}}
\end{figure}

We show in Fig. \ref{fig:dlnfdxdY_ave_ws} the numerical results for
$\left\langle d\ln f(Y,x)/dYdx\right\rangle $ applying Eq. (\ref{eq:dlnfdxdY_ave})
to the WS model. We choose $b=1.2R_{A}$ in Au+Au collisions at different
collision energies. We choose $\sigma_{NN}$ as the inelastic proton-proton
cross section determined by the global fit of the experimental data
\cite{Agashe:2014kda}, whose values are listed in Table \ref{tab2}.
Also shown in Table \ref{tab2} are the values of $\left\langle d\ln f(Y,x)/dYdx\right\rangle $
at $Y=0$ in Au+Au collisions with the HS and WS distribution. 

\begin{table}[H]
\centering\caption{The inelastic nucleon-nucleon cross section $\sigma_{NN}$ at different
collisions energies (first two rows). The numerical results of $\left\langle d\ln f(Y,x)/dYdx\right\rangle $
at $Y=0$ in Au+Au collisions at different collisions energies (last
two rows). The impact parameter is set to $b=1.2R_{A}$ corresponding
to 20-50\% centrality in experiments. \label{tab2}}
\begin{tabular}{|c|c|c|c|c|c|c|c|c|c|c|}
\hline 
$\sqrt{s_{NN}}$(GeV)  & 200  & 62.4  & 54.4  & 39  & 27  & 19.6  & 14.5  & 11.5  & 9.2  & 7.7 \tabularnewline
\hline 
$\sigma_{NN}$(mb)  & 42.0  & 36.3  & 35.2  & 33.6  & 32.8  & 32.3  & 31.8  & 31.4  & 30.9  & 30.6\tabularnewline
\hline 
$\left\langle \frac{d\ln f(Y,x)}{dYdx}\right\rangle _{\mathrm{HS}}$  & 0.0471  & 0.0602  & 0.0622  & 0.0678  & 0.0753  & 0.0833  & 0.0925  & 0.101  & 0.111  & 0.121\tabularnewline
\hline 
$\left\langle \frac{d\ln f(Y,x)}{dYdx}\right\rangle _{\mathrm{WS}}$  & 0.0374  & 0.0460  & 0.0472  & 0.0507  & 0.0558  & 0.0614  & 0.0678  & 0.0739  & 0.0809  & 0.0876\tabularnewline
\hline 
\end{tabular}
\end{table}

Similar to the results of the HS model, $\left\langle d\ln f(Y,x)/dYdx\right\rangle $
at $Y=0$ increases with decreasing collision energies, and it is
a slowly increasing function of $Y$. There are also some differences
between the WS and HS results. First, due to the smooth function in
the WS model at the edge of the nucleus, the rapidity dependence of
$\left\langle d\ln f(Y,x)/dYdx\right\rangle $ in the WS model is
slightly weaker than that in the HS model. Second, besides the explicit
collision energy dependence of $Y_{L}$, $\sigma_{NN}$ also depends
on the collision energy and enters the thickness function via Eq.
(\ref{eq:rho_AB_ws}), therefore the increase of $\left\langle d\ln f(Y,x)/dYdx\right\rangle $
at $Y=0$ in the WS model with the decreasing collision energy is
slightly slower than in the HS model. 

\begin{figure}
\centering\includegraphics[scale=0.5]{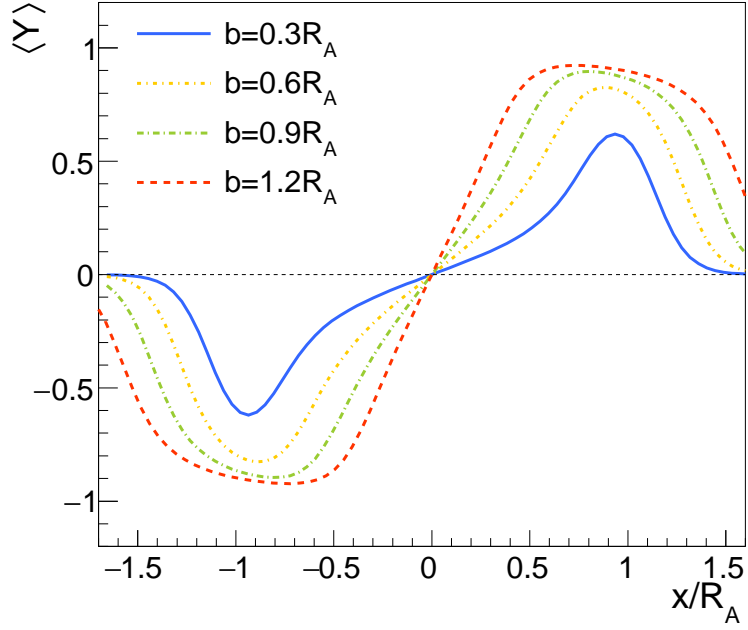}

\caption{The average rapidity in the full rapidity range for Au+Au collisions
$\sqrt{s_{NN}}=200$ GeV as a function of $x$ from Eq. (\ref{eq:ycx-1})
in the WS model. \label{fig:avY-ws}}
\end{figure}

The numerical result for the average rapidity in the full rapidity
range from Eq. (\ref{eq:ycx-1}) is shown in Fig. \ref{fig:avY-ws}
which is consistent with Fig. 5 of Ref. \cite{Gao:2007bc}. The difference
between Fig. \ref{fig:avY-ws} with the WS distribution and Fig. \ref{fig:avY}
with the HS distribution is that in the edge of the overlapping region
of two nuclei $\left\langle Y\right\rangle $ with the WS distribution
is smoothly vanishing far outside the overlapping region but $\left\langle Y\right\rangle $
with the HS distribution is discontinuous at the boundary. 

\section{Polarization induced by orbital angular momentum }

As proposed in \cite{Gao:2007bc}, the OAM in peripheral collisions
of two nuclei can induce hadron polarization. Here we assume that
the polarization is proportional the local OAM 
\begin{align}
P_{q}\left(Y\right) & =\alpha(Y)\left\langle L_{y}\right\rangle \nonumber \\
 & =-\alpha(Y)(\Delta x)^{2}\frac{\Delta_{Y}^{2}}{24}\left\langle p_{T}\right\rangle \left\langle \frac{d\ln f(Y,x)}{dYdx}\right\rangle ,\label{eq:PqY}
\end{align}
where we have replaced $d\ln f(Y,x)/dYdx$ in Eq. (\ref{eq:Ly}) with
its average in Eq. (\ref{eq:dlnfdxdY_ave}), and $\alpha(Y)$ is a
rapidity-dependent coefficient. The minus sign means that the polarization
is along $-y$ direction. We define a parameter $\kappa(Y)=\alpha(Y)(\Delta x)^{2}\Delta_{Y}^{2}$
as a function of the rapidity $Y$. Note that $\Delta x$, $\Delta_{Y}$
and $\left\langle p_{T}\right\rangle $ can also depend on $Y$ in
principle.

The global polarization of $\Lambda$ hyperons at mid-rapidity has
been measured in the STAR experiment by which the parameters in Eq.
(\ref{eq:PqY}) can be determined. At mid-rapidity $Y=0$, Eq. (\ref{eq:PqY})
becomes 
\begin{equation}
P_{q}(Y=0)=-\frac{1}{24}\kappa_{0}\left\langle p_{T}\right\rangle \left\langle \frac{d\ln f(Y,x)}{dYdx}\right\rangle _{Y=0},\label{eq:PqY0}
\end{equation}
where we have combined three parameters into one $\kappa_{0}\equiv\kappa(0)$.
The average transverse momentum $\left\langle p_{T}\right\rangle $
can be determined by kaon data available at some collision energies
and an interpolation is made for other collision energies. Results
of $\left\langle d\ln f(Y,x)/dYdx\right\rangle _{Y=0}$ are already
shown in Table \ref{tab2}.

\begin{figure}[H]
\centering\includegraphics[scale=0.4]{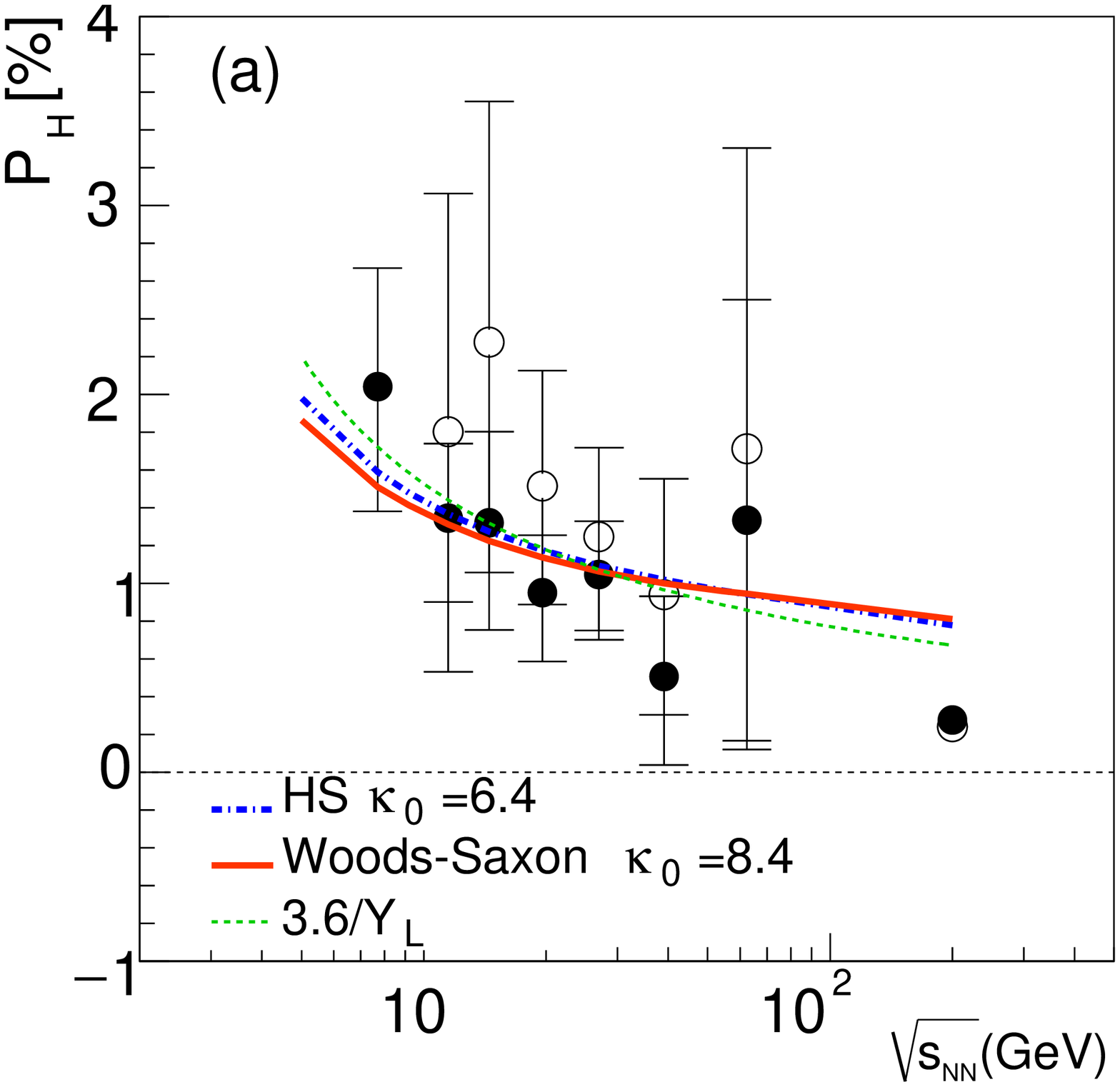}\includegraphics[scale=0.4]{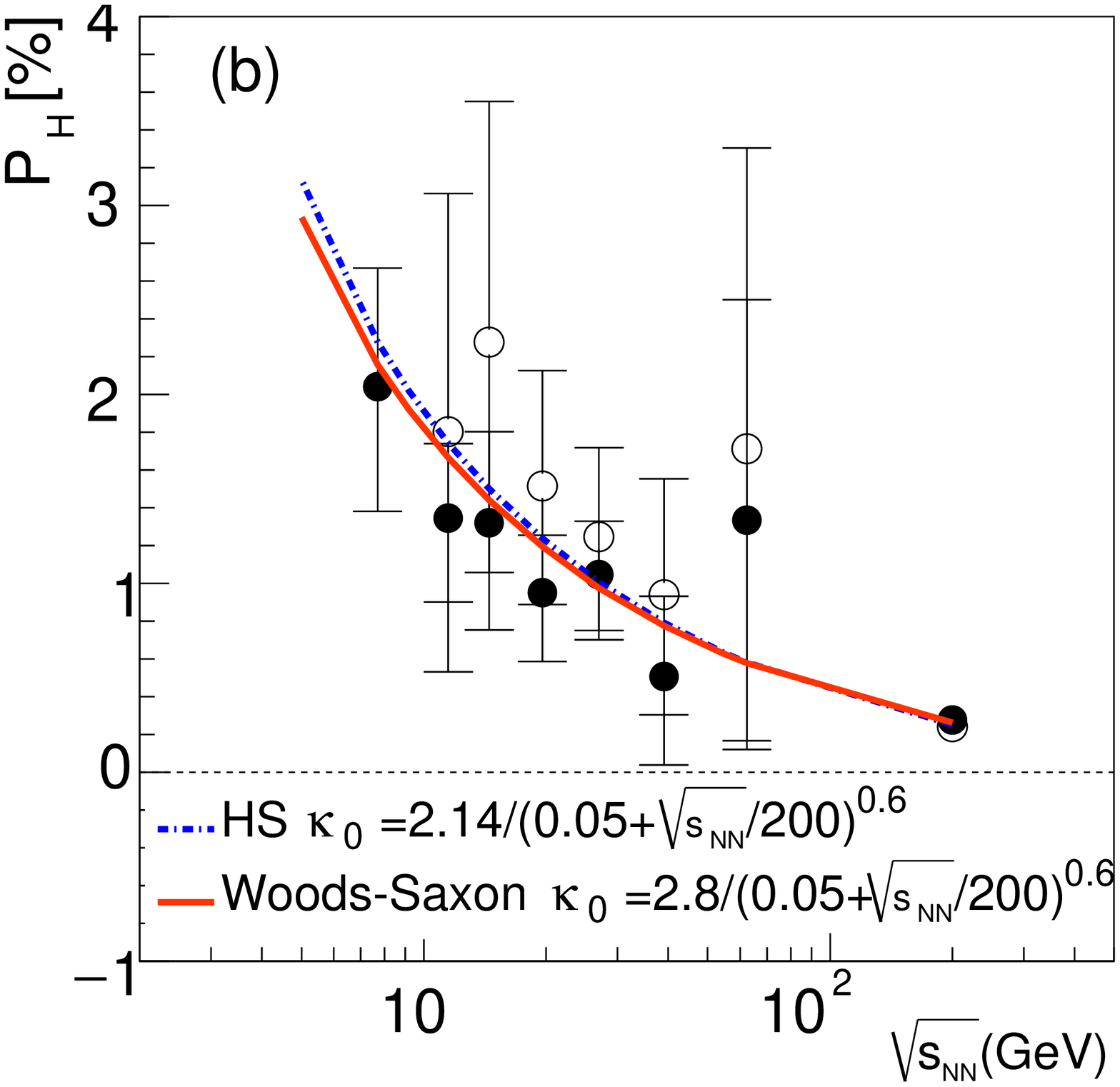}\caption{The global polarization of $\Lambda$ and $\bar{\Lambda}$ at mid-rapidity
in the HS and WS model in Au+Au collisions at different collision
energies. The impact parameter is set to $b=1.2R_{A}$ corresponding
to the centrality 20-50\%. The STAR data are represented by the solid
($\Lambda$) and open ($\bar{\Lambda}$) circle \cite{STAR:2017ckg,Adam:2018ivw}.
(a) $\kappa_{0}$ is a contant of the collision energy. (b) $\kappa_{0}$
depends on the collision energy. \label{fig:Pq_Y0}}
\end{figure}

As our first option, we assume that the parameter $\kappa_{0}$ is
a constant of the collision energy whose value is chosen to describe
via Eq. (\ref{eq:PqY0}) the polarization data in the energy range
7.7-62.4 GeV. The results are shown in the left panel of Fig. \ref{fig:Pq_Y0}.
We see that the collision energy dependence of $P_{q}(Y=0)$ in HS
with $\kappa_{0}=6.4$ and that in WS with $\kappa_{0}=8.4$ are roughly
consistent with the polarization data in the energy range 7.7-62.4
GeV. But the fitting curves are larger than the data of 200 GeV. Interestingly
we find that the energy dependence of our results in both HS and WS
model can be well fitted by $\sim1/Y_{L}$ (dashed line). As our second
option, we use the energy dependent $\kappa_{0}$ to fit the data
in the energy range 7.7-62.4 GeV including 200 GeV. The results are
shown in the right panel of Fig. \ref{fig:Pq_Y0}.

Once the values of $\kappa_{0}$ are constrained by the polarization
data at mid-rapidity, we can predict the polarization of $\Lambda$
hyperons at larger rapidity. In our prediction we take the WS model
and the energy dependent $\kappa_{0}$ as determined in the right
panel of Fig. \ref{fig:Pq_Y0}. Since we do not know the rapidity
dependence of $\kappa(Y)$ and $\left\langle p_{T}\right\rangle $
we will consider three cases and make corresponding prediction for
the rapidity dependence of the global polarization: (a) Both $\kappa(Y)=\kappa_{0}$
and $\left\langle p_{T}\right\rangle $ are taken to be constants
in rapidity. Therefore the rapidity dependence of the global polarization
is solely from $\left\langle d\ln f(Y,x)/dYdx\right\rangle $. (b)
Only $\kappa(Y)=\kappa_{0}$ is assumed to be constant in rapidity
while $\left\langle p_{T}\right\rangle $ depends on the rapidity.
The mid-rapidity values of $\left\langle p_{T}\right\rangle $ are
taken from the kaon data in Au+Au collisions in the collision energy
range 7.7-200 GeV \cite{Abelev:2008ab,Adamczyk:2017iwn}. The rapidity
dependence of $\left\langle p_{T}\right\rangle $ is given by fitting
the kaon data in Au+Au collisions at 62.4 GeV and 200 GeV \cite{Arsene:2009jg,Bearden:2004yx}.
(c) A rapidity constant $\left\langle p_{T}\right\rangle $ is adopted
which takes its value at mid-rapidity at each energy, but the rapidity
dependence of $\kappa(Y)$ is assumed to take the form in Eq. (\ref{eq:kapp-y}).
(d) Both $\kappa(Y)$ and $\left\langle p_{T}\right\rangle $ depend
on the rapidity. The rapidity dependence of $\left\langle p_{T}\right\rangle $
is the same as case (b), while the rapidity dependence of $\kappa(Y)$
is the same as case (c).

\begin{figure}[H]
\centering\includegraphics[scale=0.4]{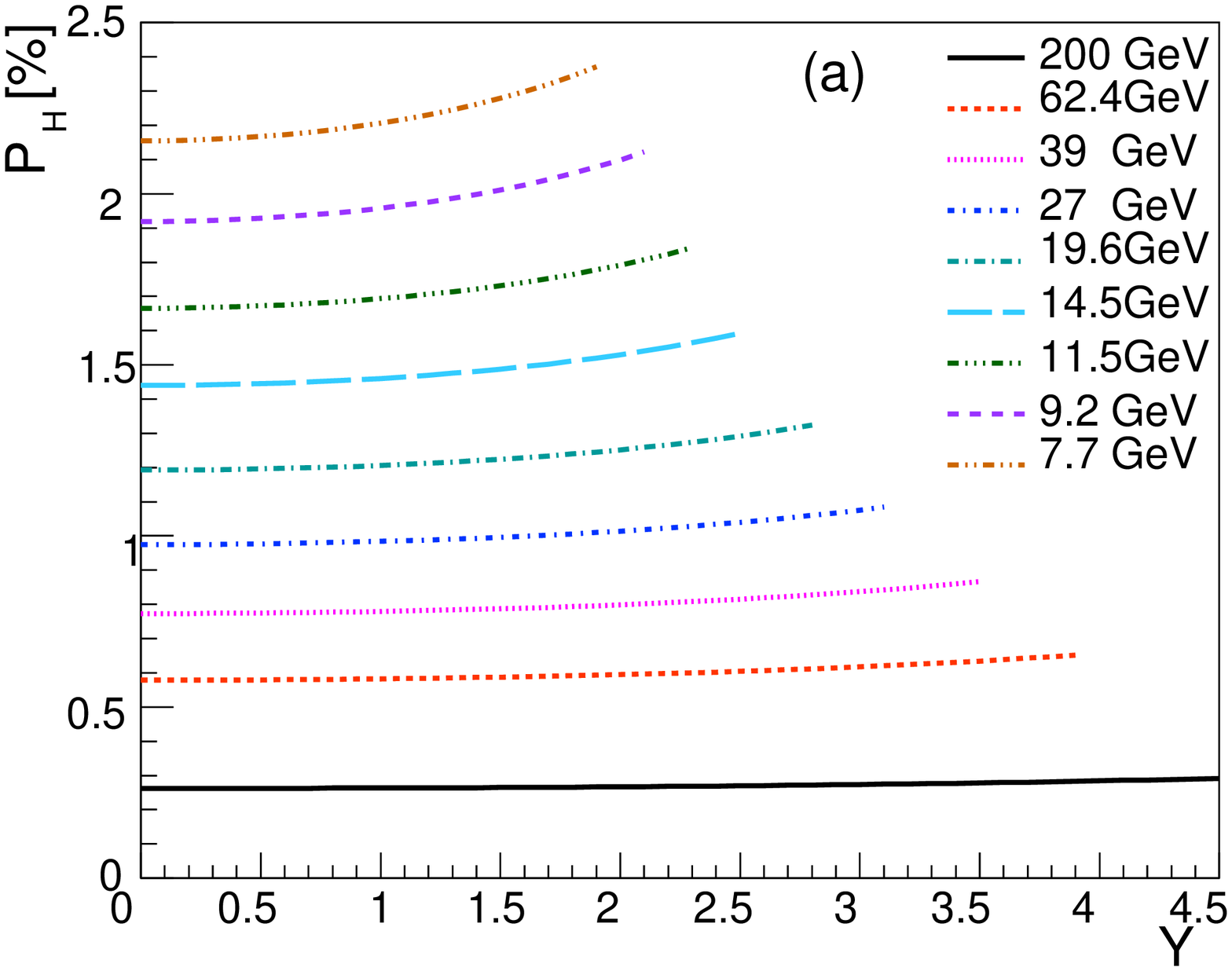}\includegraphics[scale=0.4]{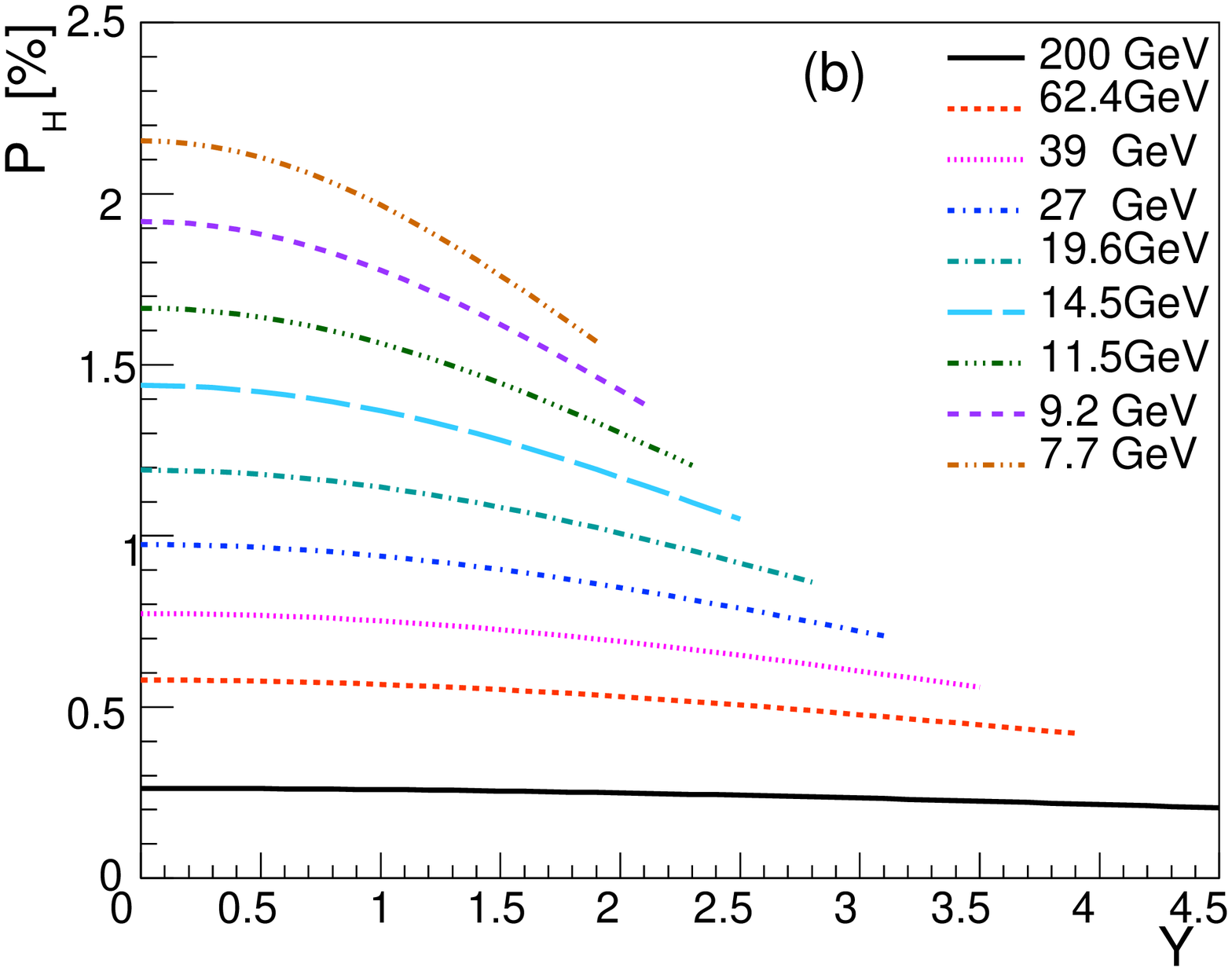}

\includegraphics[scale=0.4]{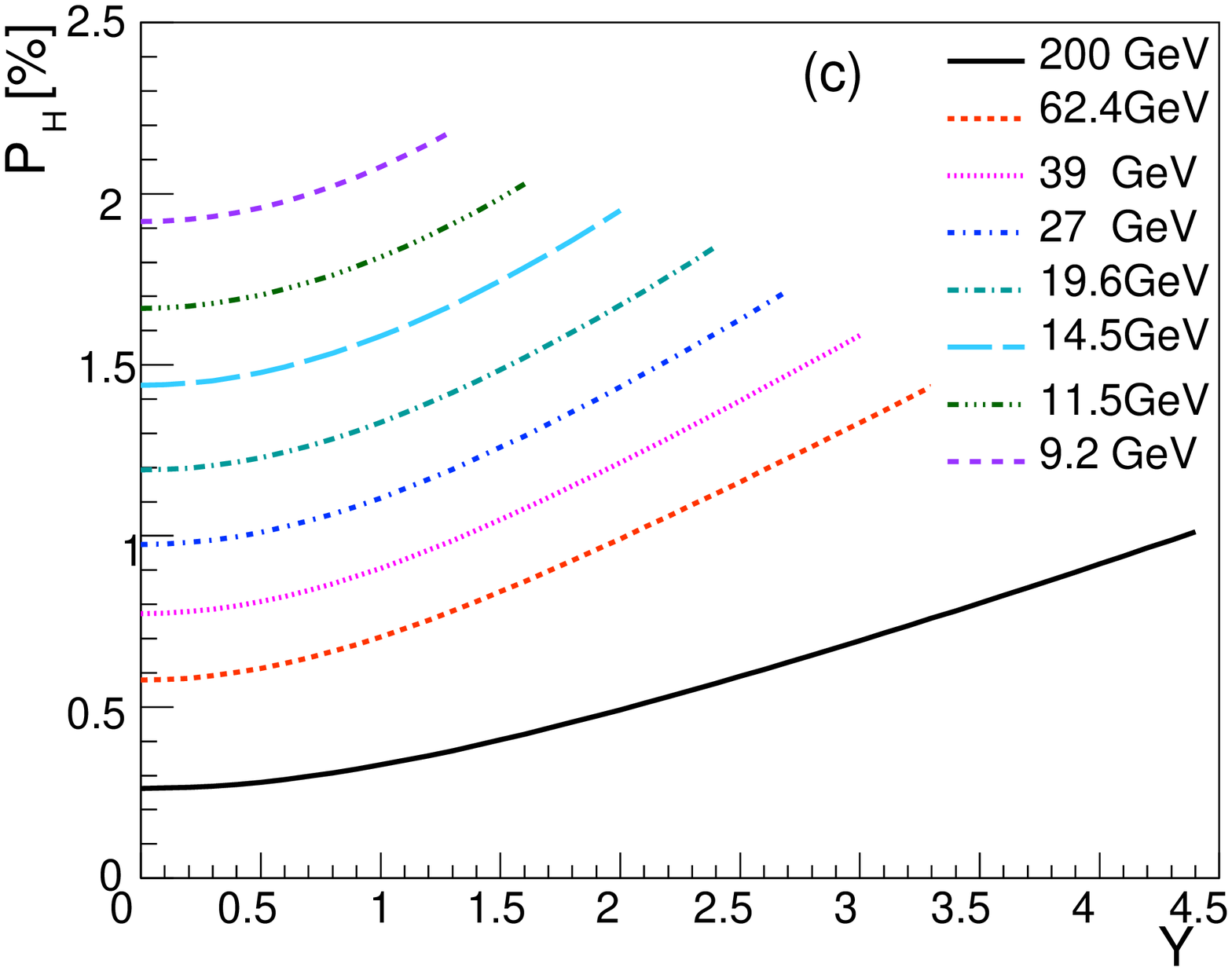}\includegraphics[scale=0.4]{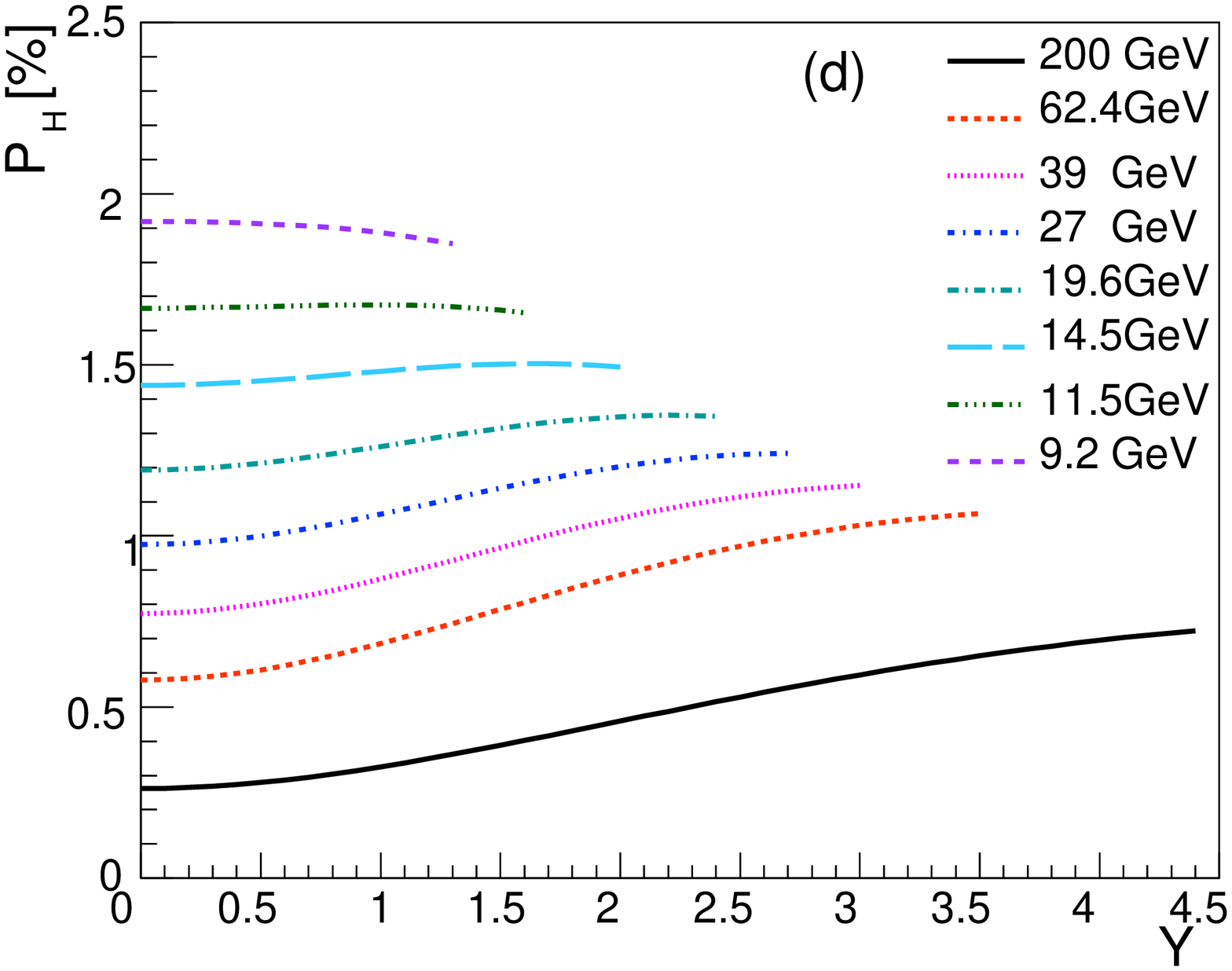}\caption{The polarization of $\Lambda$ and $\bar{\Lambda}$ in the WS model
as functions of rapidity in Au+Au collisions at different collision
energies. We use the collision energy dependent $\kappa_{0}$ as as
determined in the right panel of Fig. \ref{fig:Pq_Y0}. The impact
parameter is set to $b=1.2R_{A}$ corresponding to the centrality
20-50\%. The results in case (a), (b), (c) and (d) are presented in
panel (a), (b), (c) and (d) respectively. \label{fig:Pq_vs_Y}}
\end{figure}

Figure \ref{fig:Pq_vs_Y}(a-d) show the polarization results for case
(a-d) respectively. In Fig. \ref{fig:Pq_vs_Y}(a) for the constant
$\kappa$ and $\left\langle p_{T}\right\rangle $ in rapidity we see
that the polarization increases slighly with $Y$ at each collision
energy. The positive slope in rapidity (the increase rate of the polarization
per unit rapidity) decreases with the collision energy. Figure \ref{fig:Pq_vs_Y}(b)
shows the polarization results for the constant $\kappa$ and rapidity
dependent $\left\langle p_{T}\right\rangle $. We see that the polarization
decreases with $Y$ at each collision energy. At lower energies the
decreasing slope (the absolute value of the slope) is larger than
that at higher energies. At 200 GeV the polarization is almost a constant
of rapidity. In Fig. \ref{fig:Pq_vs_Y}(c) for the rapidity dependent
$\kappa$ and constant $\left\langle p_{T}\right\rangle $ in rapidity
we see that the polarization increases with rapidity. The lower the
collision energy the larger the increasing slope is. The increase
trend is stronger than in case (a) and (d) at the same collision energy.
Figure \ref{fig:Pq_vs_Y}(d) shows the polarization results for the
rapidity dependent $\kappa$ and $\left\langle p_{T}\right\rangle $.
At high energies the polarizations increase weakly with rapidity while
they are almost contants of rapidity at low energies. Note that the
form of $\kappa(Y)$ in Eq. (\ref{eq:kapp-y}) used in case (c) and
(d) is valid for $\mu_{B}\lesssim0.45$ GeV, however, it is beyond
such a $\mu_{B}$ limit at 7.7 GeV. Therefore the curves of 7.7 GeV
in Figs. \ref{fig:Pq_vs_Y}(c,d) are not shown since they are not
reliable. 

\section{Summary}

We propose a geometric model for the hadron polarization in heavy
ion collisions with an emphasis on the rapidity dependence. It is
based on the model of Brodsky, Gunion, and Kuhn and that of the Bjorken
scaling \cite{Gao:2007bc}. The starting point is the hadron rapidity
distribution $d^{3}N_{AA}/d^{2}\mathbf{r}_{T}dY$ as a function of
the transverse position $\mathbf{r}_{T}$ in the overlapping region
of colliding nuclei and the rapidity $Y$. We use the hard sphere
and Woods-Saxon model for the nuclear density distribution from which
the thickness function is obtained. The rapidity distribution $d^{3}N_{AA}/dxdY$
depending on the in-plane position $x$ in the overlapping region
can be derived from $d^{3}N_{AA}/d^{2}\mathbf{r}_{T}dY$ by integration
over the out-plane position $y$. The average rapidity of hadrons
can be obtained from the normalized distribution $d^{3}N_{AA}/dxdY$
or $f(Y,x)$. The collective logitudinal momentum $\left\langle p_{z}\right\rangle $
is proportional to $d\ln f(Y,x)/dY$. Then the average local orbital
angular momentum $\left\langle L_{y}\right\rangle $ is proportional
to the average of $d\ln f(Y,x)/dxdY$ over $x$, which is a function
of $Y$. The hadron polarization is assumed to be proportional to
$\alpha(Y)\left\langle L_{y}\right\rangle $ with $\alpha(Y)$ is
an unknown rapidity function representing a transfer coefficient from
the orbital angular momentum in the initial state to the hadron polarization
in the final state. There are two parameters in our model that can
have rapidity dependence. These parameters at mid-rapidity can be
constrained by the polarization data of $\Lambda$ and $\bar{\Lambda}$.
Finally we make predictions for the rapidity dependence of the hadron
polarization in the collision energy range 7.7-200 GeV by taking a
few assumed forms of the parameters. The predictions can be tested
by future experiments. 
\begin{acknowledgments}
The authors thanks C.M. Ko and J.Y. Jia for insightful discussions.
ZTL and QW are supported in part by the National Natural Science Foundation
of China (NSFC) under Grant No. 11890713 and No. 11535012.
\end{acknowledgments}

\appendix

\section{Derivation of $d\ln f(Y,x)/dYdx$ for hard sphere distribution}

\label{sec:derivation}In this Appendix we will derive $d\ln f(Y,x)/dYdx$
in the HS model for the nuclear density distribution from Eq. (\ref{eq:dnaa-dxdY})
and (\ref{eq:fyx-hs}). The definition of $C_{1}^{\pm}$ and $C_{2}^{\pm}$
are 
\begin{align}
C_{1}^{\pm} & =\int_{0}^{y_{m}}dy\left[R_{A}^{2}-(x\mp b/2)^{2}-y^{2}\right]^{1/2},\label{eq:c1_define}\\
C_{2}^{\pm} & =\int_{0}^{y_{m}}dy\left[R_{A}^{2}-(x\mp b/2)^{2}-y^{2}\right]^{-1/2},\label{eq:c2_define}
\end{align}
where $y_{m}$ is the maximum of $y$ at a fixed $x$ 
\begin{equation}
y_{m}=\left[R_{A}^{2}-(|x|+b/2)^{2}\right]^{1/2}.\label{eq:ym}
\end{equation}
The analytical expressions of $C_{1}^{\pm}$ are 
\begin{equation}
C_{1}^{+}\left(x\right)=\begin{cases}
\frac{1}{2}\sqrt{R_{A}^{2}-\left(x+b/2\right)^{2}}\sqrt{2bx}+I_{-}(x), & 0<x<R_{A}-b/2\\
\frac{\pi}{4}\left[R_{A}^{2}-\left(x-b/2\right)^{2}\right], & -(R_{A}-b/2)<x\le0
\end{cases}\label{eq:c1+}
\end{equation}
\begin{equation}
C_{1}^{-}\left(x\right)=\begin{cases}
\frac{\pi}{4}\left[R_{A}^{2}-\left(x+b/2\right)^{2}\right], & 0\le x<R_{A}-b/2\\
\frac{1}{2}\sqrt{R_{A}^{2}-\left(x-b/2\right)^{2}}\sqrt{-2bx}+I_{+}(x), & -(R_{A}-b/2)<x<0
\end{cases}\label{eq:c1-}
\end{equation}
where the function $I_{\pm}(x)$ are defined as 
\begin{equation}
I_{\pm}(x)=\frac{1}{2}\left[R_{A}^{2}-(x\pm b/2)^{2}\right]\arctan\frac{\sqrt{R_{A}^{2}-\left(x\mp b/2\right)^{2}}}{\sqrt{\mp2bx}}
\end{equation}
The analytical expressions of $C_{2}^{\pm}$ are 
\begin{align}
C_{2}^{+}\left(x\right) & =\begin{cases}
\arctan\frac{\sqrt{R_{A}^{2}-(x+b/2)^{2}}}{\sqrt{2bx}}, & 0<x<R_{A}-b/2\\
\frac{\pi}{2}, & -(R_{A}-b/2)<x\le0
\end{cases}\label{eq:c2+}
\end{align}
\begin{align}
C_{2}^{-}\left(x\right) & =\begin{cases}
\frac{\pi}{2}, & 0\le x<R_{A}-b/2\\
\arctan\frac{\sqrt{R_{A}^{2}-(x-b/2)^{2}}}{\sqrt{-2bx}}, & -(R_{A}-b/2)<x<0
\end{cases}\label{eq:c2-}
\end{align}
In terms of $C_{1}^{\pm}$ and $C_{2}^{\pm}$, we obtain the derivative
of $\ln f(Y,x)$ with respect to $Y$ as 
\begin{equation}
\frac{d\ln f(Y,x)}{dY}=\frac{d\ln(dN_{\mathrm{pp}}/dY)}{dY}+\frac{1}{Y}-\frac{1}{Y}\left\{ 1+\frac{Y}{Y_{L}}\cdot\frac{C_{1}^{+}-C_{1}^{-}}{C_{1}^{+}+C_{1}^{-}}\right\} ^{-1},
\end{equation}
and then the derivative of $d\ln f(Y,x)/dY$ with respect to $x$
as 
\begin{eqnarray}
\frac{d\ln f(Y,x)}{dYdx} & = & \frac{1}{Y_{L}}\left[1+\frac{Y}{Y_{L}}\cdot\frac{C_{1}^{+}-C_{1}^{-}}{C_{1}^{+}+C_{1}^{-}}\right]^{-2}\left\{ -\frac{x(C_{2}^{+}-C_{2}^{-})-(b/2)(C_{2}^{+}+C_{2}^{-})}{(C_{1}^{+}+C_{1}^{-})}\right.\nonumber \\
 &  & +\frac{(C_{1}^{+}-C_{1}^{-})\left[x(C_{2}^{+}+C_{2}^{-})-(b/2)(C_{2}^{+}-C_{2}^{-})\right]}{(C_{1}^{+}+C_{1}^{-})^{2}}\nonumber \\
 &  & -2(2b|x|)^{1/2}(|x|+b/2)\left[R_{A}^{2}-(|x|+b/2)^{2}\right]^{-1/2}\nonumber \\
 &  & \times\left.\frac{C_{1}^{+}\theta(-x)+C_{1}^{-}\theta(x)}{(C_{1}^{+}+C_{1}^{-})^{2}}\right\} ,\label{eq:dlnfdydx}
\end{eqnarray}
where we have used 
\begin{eqnarray}
\frac{d}{dx}\int_{0}^{a(x)}dyb(x,y) & = & \int_{0}^{a(x)}dy\frac{\partial b(x,y)}{\partial x}+b(x,a(x))\frac{da(x)}{dx}.
\end{eqnarray}

\section{Rapidity dependence of $\kappa(Y)$ in case (c) and (d)}

There is a similarity in hadron production between the case at forward
rapidity but high collision energy and that at central rapidity but
lower collision energy. Here the baryon number density or baryon chemical
potential is the relevant physical quantity. Therefore, if we neglect
the system size effect, we can make an ansatz for $\kappa(Y)$ based
on this similarity.

By fitting the data in Fig. 8(b) we find the following energy behavior
of $\kappa_{0}\equiv\kappa(Y=0)$, 
\begin{equation}
\kappa(Y=0)=\frac{2.8}{\left(0.05+\sqrt{s_{NN}}/200\right)^{0.6}}.
\end{equation}
The collision energy dependence of the baryon chemical potential at
mid-rapidity can be given by \cite{Andronic:2017pug}
\begin{equation}
\mu_{B}(Y=0)=\frac{1.3075}{1+0.288\sqrt{s_{NN}}}\:\textrm{GeV}.
\end{equation}
We can solve $\sqrt{s_{NN}}$ as a function of $\mu_{B}(Y=0)$ and
obtain

\begin{equation}
\kappa(Y=0)=\frac{2.8}{\left(0.05+\left(1.3075/\mu_{B}(Y=0)-1\right)/57.6\right)^{0.6}}.
\end{equation}
within the range $\mu_{B}\lesssim0.45$ GeV, i.e., in the collision
energy range 7.7-200 GeV. We can generalize the above expression to
other rapidity values as 
\begin{equation}
\kappa(Y)=\frac{2.8}{\left(0.05+\left(1.3075/\mu_{B}(Y)-1\right)/57.6\right)^{0.6}},\label{eq:kapp-y}
\end{equation}
by taking the following parametrization of $\mu_{B}(Y)$ \cite{Becattini:2007ci}

\begin{equation}
\mu_{B}(Y)=\mu_{B}(Y=0)+w\left(\sqrt{s_{NN}}\right)Y^{2},
\end{equation}
with the width parameter 
\begin{equation}
w\left(\sqrt{s_{NN}}\right)=0.09527-0.01594\log\sqrt{s_{NN}}.
\end{equation}

\bibliographystyle{apsrev}
\bibliography{oam}

\end{document}